\documentclass[12pt]{article}

\usepackage{graphicx}
\usepackage{color}
\usepackage{amssymb}
\usepackage{latexsym}

\font\manual=manfnt \def\dbend{\lower3.5pt\hbox{\manual\char127}}

\def\ie{{\it i.e.}}
\def\eg{{\it e.g.}}
\def\cf{{\it c.f.}}

\def\sst{\scriptscriptstyle}

\def\frac#1#2{{#1\over#2}}
\def\coeff#1#2{{\textstyle{#1\over #2}}}
\def\half{\frac12}
\def\hf{{\textstyle\half}}

\def\IR{{\mathbb R}}

\def\IS{{\mathbb S}}
\def\IZ{{\mathbb Z}}

\def\One{{1\hskip -3pt {\rm l}}}

\catcode`\@=11
\def\slash#1{\mathord{\mathpalette\c@ncel{#1}}}
\def\underrel#1\over#2{\mathrel{\mathop{\kern\z@#1}\limits_{#2}}}

\catcode`\@=12
\def\ket#1{|#1\rangle}
\def\bra#1{\langle#1|}
\def\vev#1{\langle#1\rangle}
\def\det{{\rm det}}
\def\tr{{\rm tr}}

\def\sinh{{\rm sinh}} 	
\def\cosh{{\rm cosh}} 	\def\ch{{\rm ch}}
\def\tanh{{\rm tanh}}

\def\det{{\rm det}}

\def\exp{{\rm exp}}

\def\DD{{\cal D}} 
\def\EE{{\cal E}} 
 
\def\GG{{\cal G}} 
\def\HH{{\cal H}}

\def\OO{{\cal O}} \def\CO{{\cal O}}
\def\PP{{\cal P}} 
 
\def\RR{{\cal R}} 
\def\SS{{\cal S}} 
 
\def\UU{{\cal U}}

\def\ZZ{{\cal Z}} 
\def\lam{\lambda}
\def\eps{\epsilon}
\def\vareps{\varepsilon}
\def\unlockat{\catcode`\@=11}
\def\lockat{\catcode`\@=12}
\unlockat
\def\newsec#1{\global\advance\secno by1\message{(\the\secno. #1)}
\global\subsecno=0\global\subsubsecno=0\eqnres@t\noindent
{\bf\the\secno. #1}
\writetoca{{\secsym} {#1}}\par\nobreak\medskip\nobreak}
\global\newcount\subsecno \global\subsecno=0
\def\subsec#1{\global\advance\subsecno
by1\message{(\secsym\the\subsecno. #1)}
\ifnum\lastpenalty>9000\else\bigbreak\fi\global\subsubsecno=0
\noindent{\it\secsym\the\subsecno. #1}
\writetoca{\string\quad {\secsym\the\subsecno.} {#1}}
\par\nobreak\medskip\nobreak}
\global\newcount\subsubsecno \global\subsubsecno=0
\def\subsubsec#1{\global\advance\subsubsecno by1
\message{(\secsym\the\subsecno.\the\subsubsecno. #1)}
\ifnum\lastpenalty>9000\else\bigbreak\fi
\noindent\quad{\secsym\the\subsecno.\the\subsubsecno.}{#1}
\writetoca{\string\qquad{\secsym\the\subsecno.\the\subsubsecno.}{#1}}
\par\nobreak\medskip\nobreak}
\def\subsubseclab#1{\DefWarn#1\xdef
#1{\noexpand\hyperref{}{subsubsection}%
{\secsym\the\subsecno.\the\subsubsecno}%
{\secsym\the\subsecno.\the\subsubsecno}}%
\writedef{#1\leftbracket#1}\wrlabeL{#1=#1}}
\lockat
\newcommand{\mthb}[1]{\mbox{\boldmath \(#1\)}}

\newcommand{\be}{\begin{equation}}
\newcommand{\ee}{\end{equation}}

\newcommand{\bbb}{\begin{eqnarray}}
\newcommand{\eee}{\end{eqnarray}}
\newcommand{\pref}[1]{(\ref{#1})}

\begin{document}
 

%
\def\ws{{\rm \sst WS}}
\def\alp{\alpha'}
\def\eff{{\rm eff}}
\def\mm{{\mthb\mu}}
\def\mub{\mu_{\!B}^{~}}
\def\muhat{\hat\mu}
\def\mubhat{\hat\mub}
\def\bb{{\tt b}}
\def\cc{{\tt c}}
\def\pp{{\bf p}}
\def\qq{{\bf q}}
\def\xx{{\bf x}}
\def\lz{{\tt z}}
\def\fzzt{{\rm FZZT}}
\def\zz{{\rm ZZ}}
\def\cp{{\rm\sst CP}}
\def\lpl{\ell_{\rm pl}}
\def\lstr{\ell_{\rm s}}
\def\sx{{\sst X}}
\def\sphi{{\sst\phi}}
\def\V{{\bf V}}
\def\moduli{1}
\def\strloops{2}
\def\tachwall{3}
\def\bounce{4}
\def\tiles{5}
\def\props{6}
\def\vortex{7}
\def\cubicpotl{8}
\def\phasespace{9}
\def\macroloop{10}
\def\wavefn{11}
\def\oneptfn{12}
\def\loopamps{13}
\def\annulus{14}
\def\senpotl{15}
%

 
\begin{titlepage}
\rightline{EFI-04-34}
 
\rightline{hep-th/0410136}
 
\vskip 3cm
\centerline{\Large{\bf Matrix Models and 2D String Theory}}
 
\vskip 2cm
\centerline{
Emil J. Martinec\footnote{\texttt{e-martinec@uchicago.edu}}}
\vskip 12pt
\centerline{\sl Enrico Fermi Inst. and Dept. of Physics}
\centerline{\sl University of Chicago}
\centerline{\sl 5640 S. Ellis Ave., Chicago, IL 60637, USA}
 
\vskip 2cm
 
\begin{abstract}
String theory in two-dimensional spacetime illuminates
two main threads of recent development in string theory:
(1)~Open/closed string duality, and 
(2)~Tachyon condensation.
In two dimensions, many aspects of these phenomena
can be explored in a setting where exact calculations
can be performed.  These lectures review the basic
aspects of this system.
\end{abstract}

\end{titlepage} 
 
\newpage
 
\setcounter{page}{1}

 
\section{\label{introsec}Introduction}

One of the most remarkable developments in string theory in recent
years is the idea that, in certain circumstances
(superselection sectors), it has a presentation as a large N
gauge dynamics -- gravitation is a collective phenomenon
of the gauge theory, and closed strings are represented
by loop observables of the gauge theory.  
The gauge theory in these situations provides an ansatz
for the nonperturbative {\it definition} of the theory
in that superselection sector.

By superselection sector, one means a choice of asymptotic behavior
for the low-energy fields.
A canonical example is string theory in $AdS_5\times \IS^5$
(for a review, see \cite{Aharony:1999ti}),
where the states of the theory all have a metric that
asymptotes to the anti-de Sitter metric times a round sphere,
and the self-dual five-form field strength of type IIB supergravity
carries $N$ units of flux through the $\IS^5$.
The gauge theory equivalent is maximally supersymmetric
$U(N)$ Yang-Mills theory in $D=4$ spacetime dimensions.  
The correspondence equates states of geometry and matter 
in this superselection sector with states of the gauge theory.  
Both $AdS_5\times \IS^5$
and maximally supersymmetric gauge theory possess
the same global superconformal symmetry 
($SU(2,2|4)$ in the language of supergroups),
which then organizes the state space into representations
of the superconformal algebra.
For instance, one can match the one-particle states,
and the operators that create them from the vacuum,
by their representation properties.
The operators that create and destroy strings are represented in
the gauge theory description by Wilson loops,
$\tr[\exp \,i\!\oint \!A]$, and their supersymmetric generalizations.

Upon the injection of a little energy, in gravity the
generic state is a gas of supergravitons 
(the graviton and particles related to it by supersymmetry);
if we put in a lot of energy, we expect a black hole to form.
On the gauge theory side, at low energies the excitations are
built from collections of gauge singlet operators 
(multiple `single-particle creation operators')
acting on the vacuum; at high energies, 
the gauge theory undergoes a ``deconfinement transition''
where energy is equipartitioned into all $N^2$ fields 
of the matrix field theory.  The correspondence
equates the transition from supergraviton gas to black hole 
on the geometry side, and the deconfinement transition on the 
gauge theory side \cite{Witten:1998zw}.

Indeed, this equivalence first arose via the study of black holes
carrying D-brane charge in string theory
(for a review, see \cite{Peet:2000hn}).  On the one hand,
the dynamics on the branes is described at low energies
by the lightest strings attached to the branes.  
The spacetime effective field theory, in which
these strings are the quanta, is a Yang-Mills gauge theory 
with various matter fields.  On the other hand,
the branes source a geometry in which there is an increasing
redshift of physics near the branes, as seen by asymptotic
observers.  Thus low energy also means gravitational physics
near the branes.  The gauge/gravity equivalence is the
statement that these two descriptions 
have an overlapping region of validity, namely that of
objects near the branes at low energies.
In particular, geometrical excitations of the brane
typically lead to horizon formation (`black' branes),
whose thermodynamic properties 
(\cf\ \cite{Peet:2000hn,Aharony:1999ti})
can be compared to those of the gauge theory
in the cases where they can be computed.

The loop variables describing strings in the gauge theory
representation are often cumbersome to work with,
and it remains a problem to dig out quasi-local gravitational
and other closed string physics from this exact formulation.
For instance, the local physics of the horizon and singularity
of black holes and black branes are not well-understood in the gauge
theory language (although there is some recent progress 
\cite{Fidkowski:2003nf}).
It would be useful to have a well-developed 
dictionary translating between gauge theory quantities and the 
standard perturbative formulation of string theory as a 
sum over surfaces.  Generally, we don't know how to read off
local physics beyond qualitative statements which are
dictated by symmetries (in particular, by scaling arguments)
\cite{Peet:1998wn,Banks:1998dd,Balasubramanian:1998de}.

Part of the reason that this dictionary is poorly developed
is that the correspondence is a strong/weak coupling duality.
The radius of curvature $R$ of both $AdS_5$ and $\IS^5$,
relative to the Planck scale $\lpl$ of quantum gravity,
is $N=(R/\lpl)^4$; relative to the scale $\lstr$
set by the string tension, it is $g_{\rm\sst YM}^2N=(R/\lstr)^4$.
Thus for the spacetime to have a conventional interpretation as
a geometry well-approximated by classical Einstein gravity,
we should work in the gauge theory at both large $N$ 
and large effective ('t Hooft) coupling strength $g_{\rm\sst YM}^2N$.
Thus when stringy and quantum gravity fluctuations are suppressed,
the gauge theory description is strongly coupled; and when
the gauge theory is perturbative, the geometry has unsuppressed
quantum and stringy effects.

Often in physics, useful information can be gathered by consideration
of low-dimensional model systems, which hopefully retain essential features
of dynamics, while simplified kinematics renders precise analysis possible.
If one or another side of the duality is exactly solvable,
then we can bypass the difficulty of strong/weak duality.

String theory in two spacetime dimensions
provides just such an example of the
gauge/gravity (or rather open string/closed string) correspondence,
in which the gauge theory is an exactly solvable
random matrix model, and the worldsheet description 
of string theory involves a conformal field theory (CFT)
which has been solved by conformal bootstrap techniques.

The random matrix formulation of 2D string theory
was discovered well before the recent developments 
involving D-branes; in fact it provided some 
of the motivation for the discovery of D-branes.
The initial work on the matrix model is reviewed
extensively in \cite{Klebanov:1991qa,Ginsparg:1993is}.
The exact solution of Liouville theory was not developed
at that time, and so precise comparison with
worldsheet computations was rather limited in scope.
The development of the conformal bootstrap for Liouville
\cite{%
Dorn:1994xn,%
Zamolodchikov:1995aa,%
Teschner:1995yf,%
Fateev:2000ik,%
Teschner:2000md,%
Zamolodchikov:2001ah},
reviewed in \cite{Teschner:2001rv,Nakayama:2004vk},
took place in the following decade,
while much of string theory research was focussed
on gauge/gravity equivalence.  It has only been
in the last year or so that these various threads
of research have been woven together
\cite{%
McGreevy:2003kb,%
Martinec:2003ka,%
Klebanov:2003km,%
Takayanagi:2003sm,%
Douglas:2003up}.

Our goal in these lectures will be to provide a
self-contained overview this system,
giving an introduction to the matrix model of 2D string theory,
as well as the CFT techniques used to calculate
the corresponding perturbative string amplitudes.  
We will then illustrate the map between 
these two presentations of 2D string theory.

Along the way, we will encounter a second major theme in 
recent string research~-- the subject of {\it tachyon condensation}
(for reviews, see \cite{Sen:1999mg,Martinec:2002tz}.
A tachyon is simply terminology for an instability,
a perturbation which grows exponentially
instead of undergoing small oscillations.
Loosely speaking, in the `effective potential' of string theory, 
one has chosen to start the world at a local maximum of some component
of the `string field'.  By condensing this mode,
one learns about the topography and topology of this
effective potential, and thus about the vacuum structure
of string theory.

Much effort has gone into understanding the tachyons associated
to the decay of unstable collections of D-branes in string theory.
Here the unstable mode or modes are (open) strings
attached to the brane or branes.  For example, when one has
a brane and an anti-brane, the initial stages of their mutual
annihilation is described by the condensation of the lightest
(in this case, tachyonic) open string stretching between
the brane and the anti-brane.  Eventually the brane decays completely
into (closed) string radiation.  One might wonder
whether there is a region of overlapping validity
of the two descriptions, just as in the 
gauge/gravity (open string/closed string) correspondences
described above.  We will see evidence that this is the case in 2D
string theory. 
The random matrix presentation of 2D string theory 
was first introduced as an alternative way to describe
the worldsheets of closed 2D strings, yet the evidence suggests
that it is in fact a description of the open
string tachyon condensate on unstable D-particles.

The lectures are aimed at a broad audience; along the way,
many ideas familiar to the practicing string theorist
are summarized in order to make the presentation
as self-contained as possible.  We begin with 
a brief overview of perturbative string theory as a way of
introducing our primary subject, which is string theory
in two-dimensional backgrounds.  


\section{\label{strth}An overview of string theory}

String theory is a generalization of particle dynamics.%
\footnote{For a more detailed introduction,
the reader may consult the texts \cite{Green:1987sp,Polchinski:1998rq}.}
The sum over random paths gives a representation of the 
particle propagator
\bbb
  G(x,x') &=& \bra{x'}\coeff{i}{\partial^2-m^2}\ket{x}
\nonumber\\
	&=& \int_0^\infty \!dT \bra{x'}e^{iT(\partial^2-m^2)}\ket{x} 
\nonumber\\
	&=& \int_{\sst X(0)=x \atop \sst X(1)=x'} 
		\frac{\DD g\,\DD X}{{\rm Diff}}
	 \;\exp\Bigl[{i\int_0^1 \!dt\sqrt g 
		[g^{tt}\partial_t X^2 + m^2]}\Bigr]\ .
\label{ptclprop}
\eee
In the second line, the use of the proper time ({\it Schwinger})
parametrization turns the evaluation of the propagator into
a quantum mechanics problem, which can be recast
as a path integral given by the last line.
The introduction of intrinsic worldline gravity via the
worldline metric $g_{tt}$, while not essential,
is useful for the generalization to string theory.
The worldline metric $g_{tt}$ acts as a Lagrange multiplier
that enforces the constraint
\be
  T_{tt} = (\partial_t X)^2 - g_{tt} m^2 = 0\ ;
\label{ptclconst}
\ee
apart from this constraint, the dynamics of worldline gravity is trivial.
Indeed, we can fix a gauge $g_{tt}=T$,%
\footnote{Infinitesimal reparametrizations 
$\delta g_{tt}=\partial_t v_t$ must fix
the endpoints of the parameter space, \ie\
$v_t=0$ at $t=0,1$.  A consequence is that the constant
mode of $g_{tt}$ cannot be gauged away, 
but everything else in $g_{tt}$ can be fixed, 
allowing the choice $g_{tt}=T$;
the integral over metrics modulo reparametrizations thus
reduces to the ordinary integral $\int\! dT$.
An anologous phenomenon occurs in the string path integral;
the analogues of the parameter $T$ are the {\it moduli}
of the string worldsheet.}
and after rescaling $\tau=T t$, equation \pref{ptclprop} 
boils down to the standard path integral representation
\be
  G(x,x') = \int_{\sst X(0)=x \atop \sst X(T)=x'} \DD X\;
	\exp\Bigl[i\int_0^T \! d\tau\,[(\partial_\tau X)^2+m^2]\Bigr]\ .
\label{simpleprop}
\ee

We can generalize this construction in several ways.
For instance, we can put the particle in a curved spacetime
with metric $G_{\mu\nu}(X)$,
and in a background potential $V(X)$ that generalizes the constant $m^2$;%
\footnote{Note that $m^2$ can be thought of as
a worldline cosmological constant.}
also, we can couple a charged particle to a background
electromagnetic field specified by the vector potential
$A_\mu(X)$.  The effect is to replace the free particle
action in \pref{ptclprop}
by a generalized `worldline nonlinear sigma model'
\be
  \SS_{\rm worldline} = \int\!dt \Bigl[
	\sqrt gg^{tt}G_{\mu\nu}(X)\partial_t X^\mu\partial_t X^\nu
	+ A_\mu(X)\partial_t X^\mu - \sqrt g V(X) \Bigr]\ .
\label{ptclact}
\ee

String theory introduces a second generalization, replacing
the notion of dynamics of pointlike objects to that of
extended objects such as a one-dimensional string.
Perturbative string dynamics is governed by an action
which is the analogue of \pref{ptclact}
\be
  \SS_\ws = \frac{1}{4\pi\alp} \int d^2\sigma \Bigl[
	\Bigl(\sqrt g g^{ab}G_{\mu\nu}(X)+\epsilon^{ab}B_{\mu\nu}(X) \Bigr)
	\partial_a X^\mu\partial_b X^\nu
	+\alp\sqrt gR^{\sst (2)}\,\Phi(X) +\sqrt g\, V(X)\Bigr]
\label{sigmod}
\ee
where $a,b=0,1$ and $\mu,\nu=0,...,D-1$ are worldsheet and
target space indices, respectively.  The quantity $\alp=\lstr^2$
sets a length scale for the target space parametrized by $X^\mu$;
it plays the role of $\hbar$ for the generalized nonlinear sigma model
\pref{sigmod}.  The antisymmetric tensor gauge field $B_{\mu\nu}$
is the direct generalization of the vector potential $A_{\mu}$;
the former couples to the area element $dX^\mu\wedge dX^\nu$
of the two-dimensional string worldsheet in the same way that
the latter couples to the line element $dX^\mu$ of the particle
worldline.  In addition, because intrinsic curvature $R^{\sst (2)}$ 
can be non-trivial in two dimensions, one has an additional coupling 
of the curvature density to a field $\Phi$ known as the {\it string dilaton}.

The dynamical principle of the worldsheet theory is the requirement that
\be
  \vev{\cdots~ T_{ab} ~\cdots} = 0
\label{dynprinc}
\ee
in all correlation functions.  The two traceless components of these
equations play the same role as the constraint \pref{ptclconst} --
they enforce reparametrization invariance on the worldsheet.
The trace component is a requirement that the 2d QFT
of the worldsheet dynamics is {\it locally scale invariant},
\ie\ that the beta functions vanish.  For example, setting
$B_{\mu\nu}=V=0$, the conditions through one loop are
\bbb
  \beta_{G_{\mu\nu}} &=& \alp(\RR_{\mu\nu}(G)+\nabla_\mu\nabla_\nu\Phi)
	+O({\alp}^2) =0 
\nonumber\\
  \beta_\Phi &=& \frac{D-26}{6} + \alp \bigl(\coeff12\nabla^2\Phi
	+ (\nabla\Phi)^2\bigr) + O({\alp}^2) =0
\label{betafns}
\eee
where $\RR_{\mu\nu}(G)$ is the Ricci curvature of the
spacetime metric $G$, and $\nabla$ is the spacetime gradient.
Thus, a reason to be interested in string theory is that,
in contrast to the point particle, the string carries with it
the information about what spacetimes it is allowed to
propagate in -- namely, those that satisfy the Einstein
equations coupled to a scalar dilaton (and other fields,
if we had kept them nonzero).

Since the local invariances combine the reparametrization group
$\it Diff$ and the group of local scale transformations $\it Weyl$,
the appropriate replacement for \pref{ptclprop} is
\be
  \ZZ = \int \frac{\DD g\,\DD X}{\rm Diff\times Weyl}\;
	\exp[i\SS_\ws]\ .
\label{strpathint}
\ee
We can soak up the local gauge invariance by 
(locally on the worldsheet) choosing coordinates 
in which $g_{ab}=\delta_{ab}$.
One cannot choose such flat coordinates globally, however,
as one sees from the Gauss-Bonnet identity
$\int\! \sqrt g R^{\sst (2)} = 4\pi(2-2h)$.%
\footnote{It is standard practice to Wick rotate to
Euclidean worldsheets and ignore any associated
subtleties.  We will follow the standard practice here.}
Nevertheless, one can relate any metric via the symmetries
to one of a $6h-6$ parameter family
of reference metrics $\hat g_{ab}(m_r)$,
$r=1,...,6h-6$.  The parameters $m_r$
are called the {\it moduli} of the 2d surface.%
\footnote{There are a few special cases; for $h=0$,
there are no moduli, and for $h=1$ there are two moduli
(the length and twist of the propagator tube
joined to itself to make a torus).}
A simple picture of these parameters is shown in figure \moduli.

\begin{figure}[ht]
\begin{center}
\[
\mbox{\begin{picture}(320,65)(0,0)
\includegraphics[scale=.7]{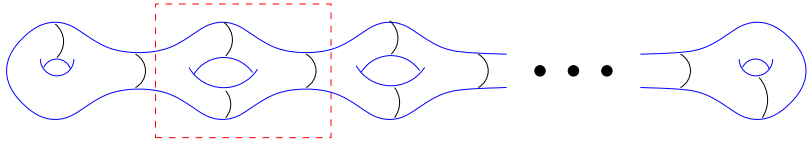}
\end{picture}}
\]
\caption{\it 
Each handle, except the end ones, contributes three 
closed string propagator tubes to the surface.
Each tube has a length and a helical twist angle.
The two end handles together contribute only three tubes, and so
the number of moduli is $6h-6$.
}
\end{center}
\end{figure}

Thus, after fixing all of the reparametrization and local scale
invariance, the integration over metrics 
$\int\!\frac{\DD g}{\rm Diff\times Weyl}$
reduces to an integration over these moduli.
The moduli are the string version of the Schwinger parametrization
of the propagator \pref{ptclprop} for a particle.
%


\section{\label{ddimstr}Strings in D-dimensional spacetime}

A simple solution to the equations \pref{betafns}
uses `conformally improved' free fields:%
\footnote{Conformally improved means that while the path integral 
\pref{strpathint} is Gaussian, so the worldsheet QFT
is free, the stress tensor is modified due to
the coupling to intrinsic curvature.}
\be
  G_{\mu\nu}=\eta_{\mu\nu}\quad,\qquad 
  B_{\mu\nu}=V=0\quad,\qquad \Phi = n_\mu X^\mu
	\qquad \Bigl(n^2 = \coeff{26-D}{6\alp}\Bigr)\quad .
\label{lindil}
\ee
The geometry seen by propagating strings is flat spacetime,
with a linear dilaton.  The dilaton slope is timelike for
$D>26$ and spacelike for $D<26$.  

Just as the perturbative series for particles is a sum
over Feynman graphs, organized
in order of increasing number of loops in the graph,
the perturbative expansion for strings is organized
by the number of handles of the corresponding sum
over worldsheets, weighted by the effective coupling
$g_{\eff}$ to the power $2h-2$ (where $h$ is the
number of handles, often called the {\it genus} of the surface).

\begin{figure}[ht]
\begin{center}
\[
\mbox{\begin{picture}(260,60)(0,0)
\includegraphics[scale=.5]{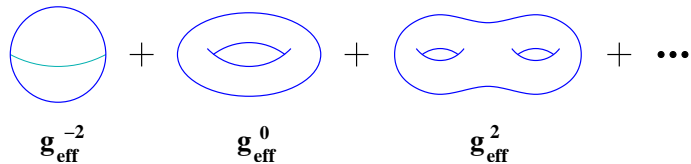}
\end{picture}}
\]
\caption{\it 
The string loop expansion.
}
\end{center}
\end{figure}

Consider string worldsheets in the vicinity of
the target space location $\hat X$.  Using
the Gauss-Bonnet identity,
the term
\be
  \frac1{4\pi}\int \sqrt g R^{\sst (2)}\,\Phi(X)
	\sim \Phi(\hat X) (2-2h)
\label{effcoup}
\ee
in the path integral over the (Euclidean) worldsheet action
identifies the effective coupling as
\be
  g_\eff = \exp[\Phi(\hat X)]\ .
\label{geff}
\ee
Thus we have strong coupling at large $\Phi=n\cdot X$,
and we have to say what happens to strings that go there.

There is also a perturbative instability of the background.
Perturbations of the spacetime background are scaling operators.
Maintaining conformal invariance at the linearized level
imposes marginality of the scaling operator.
These marginal scaling operators are known as
{\it vertex operators}.
Consider for instance adding the potential term
\be
  V(X)=\int\!\frac{d^Dk}{(2\pi)^D}\,v_k e^{ik\cdot X}
\label{tachback}
\ee
to the worldsheet action.  The scale dimension
of an individual Fourier component is determined by
its operator product with the stress tensor%
\footnote{We work in local complex coordinates
$z=\sigma^0+i\sigma^1$.}
\be
  T(z)\, e^{ik\cdot X(w)} ~~~\buildrel{z\to w}\over{\sim}~~~
	\frac{\Delta}{(z-w)^2}\, e^{ik\cdot X(w)}\ .
\label{Tope}
\ee
Using the improved stress tensor%
\footnote{As advertised, the linear dilaton determines
the conformal improvement of $T(z)=T_{zz}$ given by the second term.}
\be
  T(z) = -\coeff{1}{\alp}\partial_z X\cdot\partial_z X + 
		n\cdot\partial_z^2 X
\label{Tfree}
\ee
and evaluating the operator product expansion \pref{Tope}
via Wick contraction with the free propagator
\be
  X(z)X(w) \sim -\coeff{\alp}{2}\log|z-w|^2\ ,
\label{xprop}
\ee
one finds the scale dimension
\be
  \Delta = \coeff{\alp}{4}k^2 + \coeff{i\alp}{2} n\cdot k\quad .
\label{scaledim}
\ee
Thus the condition of linearized scale invariance
$\Delta=\bar\Delta=1$ is a mass-shell condition for $V(X)$.
This result should be no surprise -- local scale invariance
gives the equations of motion \pref{betafns} of the background,
so the linearized scale invariance condition should give the 
wave equation satisfied by small perturbations.
The mass shell condition $\Delta=1$ amounts to
\be
  (k+in)^2=-n^2-\coeff{4}{\alp}
\label{onshell}
\ee
(recall $n^2=\frac{26-D}{6\alp}$).
Thus for $D<2$, perturbations are ``massive'', and the
string background is stable.  For $D=2$, the perturbations
are ``massless'', leading to marginal stability.
Finally, for $D>2$ the perturbations are ``tachyonic'',
and the background is unstable.  The field $V(X)$ is conventionally
called the {\it string tachyon} even though 
strictly speaking that characterization only applies to $D>2$.

In the stable regime $D\le 2$, a static background condensate
$V(X)$ ``cures'' the strong coupling problem.%
\footnote{One should worry whether the conformal invariance
condition is satisfied beyond the linearized level when
one promotes the tachyon $V$ from a perturbation to a full term
in the action describing string propagation.
Fortunately, conformal invariance in 
the presence of the exponential interaction 
was demonstrated by operator methods in 
\cite{Braaten:1982yn}.
The issue is rendered moot by the construction of a conformal
bootstrap (an ansatz for the correlation functions of the exact theory) 
\cite{Dorn:1994xn,Zamolodchikov:1995aa},
which we sketch below in section \ref{wdw}.}
Let $n\cdot X=QX_1$ (recall $n^2>0$); then for $D<2$
\bbb
  V_{\rm backgd} &=& \mu\,e^{2b X_1} + \tilde\mu\, e^{2\tilde b X_1}
\nonumber\\
  {b\atop \tilde b}\Biggr\} &=& \coeff{Q}{2}\mp 
	\sqrt{(\coeff Q2)^2-\coeff 1\alp}
	= \coeff{\sqrt{26-D}\,\mp\,\sqrt{2-D}}{\sqrt{24\alp}}\ .
\label{lpotl}
\eee
(note that $\tilde b=(b\alp)^{-1}$).  
For $D=2$ one has $b=\tilde b=1/\sqrt{\alp}$, and so the 
two exponentials are not independent; rather
\be
   V_{\rm backgd}^{\sst(D=2)}
	= \mu \,X_1e^{2bX_1}+\tilde \mu\, e^{2bX_1}\ .
\label{conepotl}
\ee
The exponential barrier self-consistently keeps perturbative
string physics away from strong coupling for sufficiently
large $\mu$.

\begin{figure}[ht]
\begin{center}
\[
\mbox{\begin{picture}(280,140)(0,0)
\includegraphics[scale=.5]{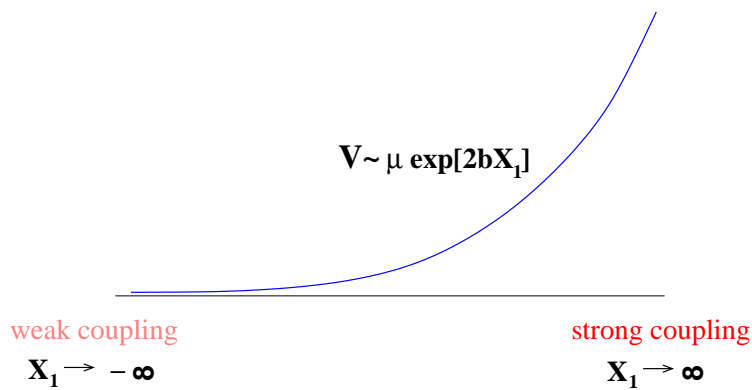}
\end{picture}}
\]
\caption{\it 
The tachyon background.
}
\end{center}
\end{figure}

For example, consider the scattering of a string tachyon of energy $E$
in $D=2$.  The string is a perturbation
$\delta V(X)=\exp[-iEX^0+ikX^1]$,
with $ik=\pm iE+Q$ the solution to the on-shell condition $\Delta=1$.
The scattering is depicted in figure \bounce.

\begin{figure}[ht]
\begin{center}
\[
\mbox{\begin{picture}(245,160)(0,0)
\includegraphics[scale=.5]{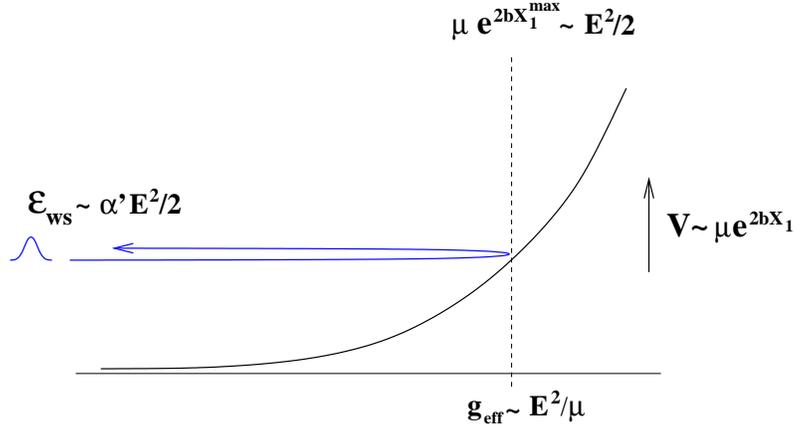}
\end{picture}}
\]
\caption{\it 
Scattering of a tachyon excitation off the tachyon background.
}
\end{center}
\end{figure}

The worldsheet energy $\EE_\ws=\alp E^2/2$ of the zero mode motion 
in $X^1$ of the string
is determined by the stress tensor $T(z)$, equation \pref{Tfree};
it is essentially the $X^1$ contribution to $\Delta$.
The turning point of the motion is determined by 
this energy to be 
\be
  V_{\rm backgd}(X_1^{\rm max})=\EE_\ws=\alp E^2/2\ .
\label{turnpt}
\ee
The effective coupling is largest at this point,
\be
  g_\eff\sim  e^{\Phi(X_1^{\rm max})}\sim {E^2}/{\mu}\ ;
\label{geffmax}
\ee
thus low energy scattering is self-consistently weakly coupled.
The effective coupling is determined by the value of the dilaton
at the turning point; we expect the scattering amplitude
to be have a perturbative series in powers of $E^2/\mu$.
Note that the high energy behavior is nonperturbative, however.

At this point we choose to relabel for $D=2$ the spacetime
coordinates
\be
  \phi\equiv X^1\quad,\qquad X\equiv X^0
\label{coordnames}
\ee
in order to conform to standard notation in the subject,
as well as to reduce the clutter of indices.
Also, we will henceforth set $\alp=1$
as a choice of units (\ie\ we measure all spacetime lengths
in ``string units'').  


\subsection{A reinterpretation of the background}

The 2d QFT%
\footnote{In an attempt to reduce confusion,
the notation 2d will be used when referring to
the string worldsheet dimension, while 2D will 
refer to two-dimensional spacetime backgrounds.}
of the ``tachyon'' background 
\be
  \SS_\ws =\frac{1}{4\pi}\int \sqrt g\Bigl[
	g^{ab}\partial_a \phi\partial_b \phi
	+ bQ \, R^{\sst (2)} \phi +\mu\, e^{2b \phi} \Bigr]
\label{xoneth}
\ee
has an alternative
interpretation in terms of worldsheet intrinsic geometry
\cite{Polyakov:1981rd},
where $e^{2b\phi}g_{ab}$ is interpreted as a 
{\it dynamical metric}, and the remaining $D-1$ fields $X$
are thought of as ``matter'' coupled to this dynamical gravity.
Let $\varphi=b \phi$; then the action becomes
\be
  \SS_\ws = \frac{1}{4\pi b^2}\int\sqrt g\Bigl[
	(\nabla\varphi)^2 + bQR^{\sst(2)}\varphi+\mu b^2\,e^{2\varphi}\Bigr]\ .
\label{liouact}
\ee
Note that $b$ plays the role of the coupling constant;
the semi-classical limit is $b\to 0$ (and thus $Q=b^{-1}+b\to b^{-1}$).
The equation of motion for $\varphi$ reads
\be
  \nabla_{\! g}^2 \,2\varphi - bQ\, R^{\sst(2)}[g] = 2\mu b^2\,e^{2\varphi}\ .
\label{lioueom}
\ee
Here $\nabla_{\! g}$ is the covariant derivative with respect to 
the intrinsic metric $g_{ab}$.
Due to the properties of the curvature under local rescaling,
\be
  \nabla_{\! g}^2 \,2\varphi-R^{\sst(2)}[g]
	=-e^{2\varphi}R^{\sst(2)}[e^{2\varphi}g]\ ,
\label{curvident}
\ee
the combination on the left-hand side of \pref{lioueom} is,
in the semi-classical limit $b\to 0$, 
just the curvature of the dynamical metric $e^{2\varphi}g_{ab}$.
The equation of motion can be written as the condition
for constant curvature of this dynamical metric
\be
  R^{\sst(2)}[e^{2\varphi}g]=-2\mu b^2\ ,
\label{lioueq}
\ee
known as the {\it Liouville equation}; the theory governed
by the action \pref{liouact} is the Liouville field theory.
The equation \pref{lioueom} is the appropriate quantum
generalization of the Liouville equation.
The constant on the right-hand side of \pref{lioueq}
is a cosmological constant for the 2d intrinsic fluctuating geometry.%
\footnote{The factor of $b^2$ on the right-hand side can be absorbed
into a renormalization of $\mu$, so that the semi-classical limit
is well-behaved.}
Note that $\sqrt g e^{2\varphi}$ is the ``dynamical area element'',
so that the potential term in the action \pref{liouact}
is a chemical potential for the dynamical intrinsic area of the
worldsheet.

This interpretation of the static tachyon background
in terms of fluctuating intrinsic geometry is only available
for $D\le 2$.  For $D>2$, the on-shell condition
$\Delta=1$ (equation \pref{onshell}) is not solved by
$V_{\rm backgd}=e^{2bX_1}$ for real $b$
(rather $b=\half Q\pm i\lambda$), and so $\sqrt g V_{\rm backgd}$
is not the area of a dynamical surface.

\subsection{KPZ scaling}

The fact that the dynamical metric is integrated over
yields useful information about the scaling of 
the partition and correlation functions with respect to 
the cosmological constant $\mu$, known as {\it KPZ scaling}
\cite{Knizhnik:1988ak,David:1988hj,Distler:1988jt}.
Consider the shift
$ \varphi\to \varphi+\frac\epsilon{2}$ 
in the Liouville action \pref{liouact} in genus $h$; this leads to
\be
  \SS_h(\mu)\longrightarrow \SS_h(e^\eps\mu)+(2-2h)\coeff Q{2b}\,\eps\ .
\label{actshift}
\ee
However, this constant mode of $\varphi$ is integrated over
in the Liouville partition function, and therefore
$\ZZ_h(\mu)$ must be independent of $\epsilon$.
We conclude
\be
  \ZZ_h(\mu) = \ZZ_h(e^\eps\mu)\, \exp[-(2-2h)\coeff Q{2b}\,\eps]
	~~~~\Longrightarrow~~~~ \ZZ_h(\mu) = c_h\,\mu^{(2-2h)Q/2b}\ .
\label{Zscaling}
\ee
For instance, for $D=1$ (pure Liouville gravity, with no matter)
one finds $\frac{Q}{2b}=\frac54$, and so the genus expansion
of the partition function is a series in $\mu^{-5/2}$.
For $D=2$, we have $\frac{Q}{2b}=1$, and so the partition
function is a series in $\mu^{-2}$.%
\footnote{In applying the KPZ scaling argument, one needs
to be sure that the path integral over the zero mode of $\varphi$
is convergent.  This is the case for $h>1$, where the Gauss-Bonnet
theorem tells us there is a classical solution to Liouville theory
since the mean curvature is negative; expanding around a metric $g_{ab}$
of constant negative curvature, the effective potential for $\varphi$
has a stable minimum due to the competition between
the exponential potential and the linear $R^{\sst(2)}\varphi$ term.
For $h=0,1$, this linear term is either absent or pushes the
wrong way; there is no local minimum for $\varphi$, and
the path integral over the zero mode diverges.  
In fact, one can show that this difficulty
occurs whenever the power of $\mu$ predicted by KPZ scaling
is non-negative.
In these cases, one can take enough derivatives
with respect to $\mu$ (\ie\ a correlation function
of several operators $e^{2b\varphi}$)
so that the scaling argument applies, and then recover
the partition function by integrating back up.
For $D=1$ this results in logarithmic corrections to
KPZ scaling, since $Q/2b$ is integral.}$^,$%
\footnote{The linear term in \pref{conepotl} 
leads to subtleties in the application of KPZ scaling,
see \cite{Polchinski:1990mf,Klebanov:1991qa,Ginsparg:1993is}.}

We could now pass to a discussion of correlation functions
of this 2d Liouville QFT, and their relation to the scattering
of strings.  Instead, we will suspend this
thread of development in favor of a random matrix
formulation of the same physics.  We will return
to the quantization of Liouville theory later,
when it is time to forge the link between these two approaches.


\section{Discretized surfaces and 2D string theory}

For spacetime dimension $D\le 2$, we have arrived at an interpretation
of the path integral describing string propagation
in the presence of a background tachyon condensate as a sum
over dynamical worldsheet geometries,
in the presence of $D-1$ ``matter fields''.%
\footnote{In solving the local scale invariance condition 
$\beta_\Phi=0$ for the string dilaton, $D-1$ is the contribution
of the fields other than $\varphi=X^1/b$ to the leading term,
the so-called {\it conformal anomaly} or {\it conformal central charge}
$$
  \vev{T_a^{~a}({\rm matter})}=\coeff{c_{\rm matter}}{48\pi}\,
	R^{\sst(2)}(g)
$$
with $c_{\rm matter}=D-1$ when the remaining theory is
conformal.  This contribution is then
cancelled by the Liouville QFT, together with a contribution
$-26$ from the reparametrization Faddeev-Popov ghosts.
This formally allows us to consider fractional $D$
(and even $D<0$ if we allow non-unitary matter CFT's),
through the use of interacting CFT's of appropriate $c_{\rm matter}$.}

A discrete or lattice formulation of fluctuating
worldsheet geometry can be given in terms of matrix Feynman graphs.  
Any tesselation of a surface built of regular polygons 
(see figure \tiles\ for a patch of tesselated surface)
has a dual%
\footnote{In the sense of Poincar\'e.}
double-line ``fatgraph'', also depicted
in figure \tiles.  The double lines indicate the
flow of matrix index contractions around the graph.

\begin{figure}[ht]
\begin{center}
\[
\mbox{\begin{picture}(385,125)(0,0)
\includegraphics[scale=.5]{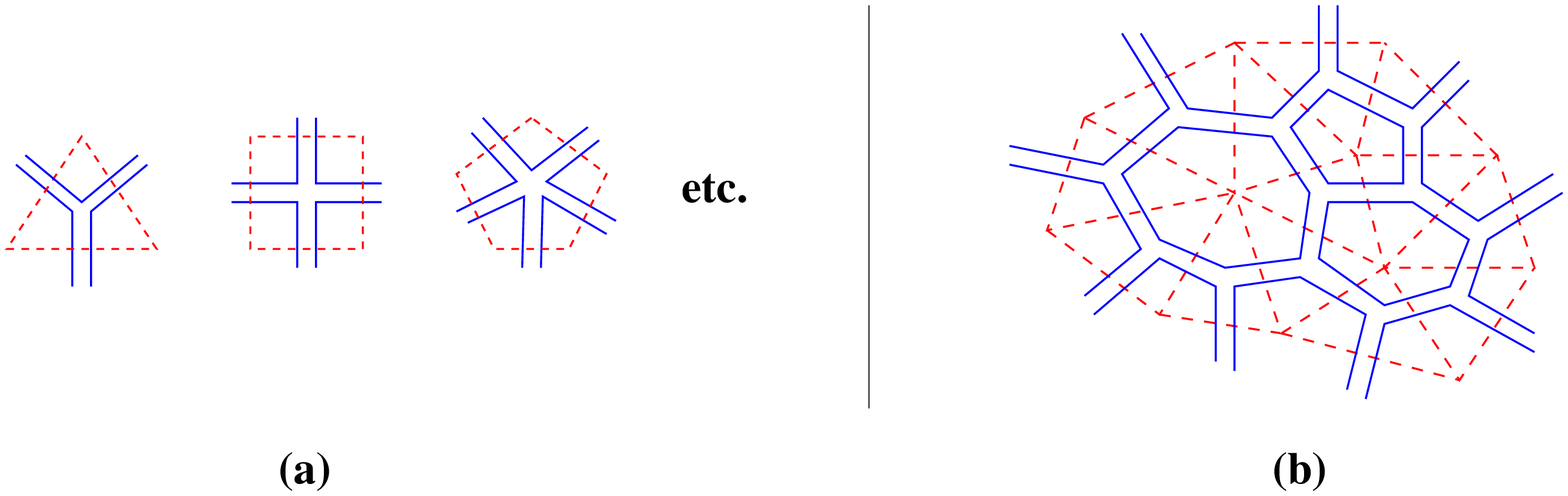}
\end{picture}}
\]
\caption{\it 
(a) Regular polygons for tiling a surface, 
with dashed red edges; 
and the dual fatgraph vertices, with solid blue dual edges.
(b) A patch of discrete surface tesselated with triangles,
and the dual fatgraph.
}
\end{center}
\end{figure}


The partition function 
\bbb
  \ZZ(g_i) &=& \int \DD^{N^2}M\, \exp[-\tr(\coeff12 g_2 M^2+ \UU(M)]
\nonumber\\
  \UU(M) &=& \coeff13 g_3 M^3 + \coeff14 g_4 M^4+\dots\quad.
\label{grgenfn}
\eee
serves as a generating function for fatgraphs,
and thereby defines an ensemble of random surfaces.  
For example, consider a surface with triangles only,
$g_{i>4}=0$.  Each face of the fatgraph gives
a factor $N$ from the trace over the index loop 
bordering the face.  Each vertex gives a factor $g_3$,
and each propagator $1/g_2$.  
The partition function 
\be
  \ZZ(g) = \sum_{V,E,F} (g_3)^V (1/g_2)^E N^F \, d(V,E,F)
\label{graphptfn}
\ee
sums over the number $d(V,E,F)$ graphs with $V$ vertices,
$E$ edges (propagators), and $F$ faces.
Using the fact that each propagator shares two vertices,
and each vertex ends three propagators, one has $2E=3V$.
The discrete version of the Gauss-Bonnet theorem
(the Euler identity) is $V-E+F=2-2h$.
The partition function is thus
\be
  \ZZ(g) =\sum_{h=0}^\infty \sum_{A}
	N^{2-2h}\Bigl(\frac{g_3 N^{1/2}}{g_2^{3/2}}\Bigr)^A \, d(h,A)
\label{thooftexpn}
\ee
where here and hereafter we write $V=A$, since the number of vertices $A$
is the discrete area of the surface.
Large $N$ thus controls the topological expansion:
$g_s^{\rm discrete}=1/N$ is the string coupling of
the discrete theory.  The cosmological constant
of the discrete theory is the free energy cost of adding area (triangles): 
$\mu_{\rm discrete} = -\log({g_3 N^{1/2}}/{g_2^{3/2}})$.


Being a lattice theory, in order to compare with the continuum
formulation of previous sections we need to take
the continuum limit of the matrix integral.  That is, we want to send the
discrete area $A$ to infinity in units of the lattice spacing
(or equivalently, send the lattice spacing to zero
for a ``typical'' surface in the ensemble).

Taking this limit amounts to balancing the suppression of 
surface area by the 2d cosmological constant 
$\mu_{\rm discrete}$ against the entropy $d(h,A)$
of large Feynman graphs (roughly, if we want to add an extra
vertex to a planar graph, there are of order $A$ places to put it).
In other words, one searches for a phase transition
or singularity in $\ZZ(g)$ where for some $g_{\rm crit}$
the partition sum is dominated by graphs with an asymptotically
large number of vertices.  Universality of this kind of critical
phenomenon is the statement that the critical point is largely 
independent of the detailed form of the matrix potential $\UU(M)$,
for instance whether the dual tesselation uses triangles
or squares in the microscopic theory
(\ie\ $M^3$ vs. $M^4$ interaction vertices in the graphical expansion).

Before discussing this phase transition, let us add in
the matter.  We wish to put discretized scalar field theory
on the random surfaces generated by the path integral over $M$.
The following modification does the job:
\be
  \ZZ = \int\DD M\, \exp\Bigl[\tr\Bigl(
	\int\! dx\int\! dx'\, \coeff12 M(x) G^{-1}(x-x')M(x')
	+\int dx\,\UU\bigl(M(x)\bigr)\Bigr)\Bigr]\ .
\label{conemm}
\ee
In the large $N$ expansion, we now have a propagator
$G(x-x')$ in the Feynman rules (rather than $g_2^{-1}={\it const.}$).
Thus, on a given graph we have a product of propagators
along the edges
\be
  \prod_{{\rm edges}}({\rm propagators})
	= \prod_{i,j\atop{\rm neighbors}} G(x_i-x_j)\ ;
\label{prodprops}
\ee
the choice $G(x-x')=\exp[-(x-x')^2/\beta]$ leads
to the discretized kinetic energy of a scalar field $X$
\be
  \prod_{i,j\atop{\rm neighbors}} G(x_i-x_j)
	= \exp\Bigl[-\frac1\beta\sum_{i,j\atop {\rm nghbrs}}(x_i-x_j)^2\Bigr]
\label{sclrKE}
\ee
which is the appropriate path integral weight for a scalar
field on the lattice.  The evaluation of the graph involves an 
integral $\prod_i\int\!dx_i$ over the location in $x$-space
of all the vertices.  In other words, we path integrate
over the discretized scalar field with the 
probability measure \pref{sclrKE}.

\begin{figure}[ht]
\begin{center}
\[
\mbox{\begin{picture}(125,125)(0,0)
\includegraphics[scale=.5]{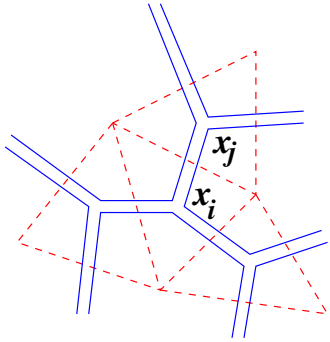}
\end{picture}}
\]
\caption{\it 
Fatgraph vertices live at points in $x$ space.
The product over propagators weights the sum over
configurations $\{x_i\}$ by a nearest-neighbor interaction
determined by the propagator $G(x_i-x_j)$.
}
\end{center}
\end{figure}

Unfortunately, the gaussian kinetic energy that leads to this
form of the propagator is not standard.
Fortunately, for $D=2$ (\ie\ one scalar matter field)
the choice $G(x-x')=\exp[-|x-x'|/\beta^{1/2}]$
turns out to be in the same universality class, and 
arises from a canonical kinetic energy for the
matrix path integral
\bbb
  G^{-1}(p)=e^{\beta p^2} &\longleftrightarrow& 
		G(x)=(\coeff{\pi}{\beta})^{\frac12}\,e^{-x^2/4\beta}
\nonumber\\
  \tilde G^{-1}(p)=1+\beta p^2 &\longleftrightarrow& 
		\tilde G(x)=\coeff{\pi}{\sqrt\beta}\,e^{-|x|/\beta^{1/2}}
\label{sameclass}
\eee
The continuum limit involves scalar field configurations 
which are slowly varying on the scale of the lattice spacing,
which is enforced by taking $\beta p^2\to 0$.
But in this limit $G^{-1}\sim \tilde G^{-1}$
and so we expect the two choices to lead to the same continuum
physics.  But in $D=2$ (\ie\ one-dimensional $x$-space),
$\tilde G(p)$ is the conventional Feynman propagator for $M$,
and so we may write
\be
  \ZZ = \int\!\DD M\,\exp\Bigl\{-\!\int\!\! dx\;\tr
	\Bigl[\frac\beta2\Bigl(\frac{dM}{dx}\Bigr)^2 \!- \UU(M)\Bigr]\Bigr\}\ ,
\label{mqm}
\ee
now with $\UU(M)=-\frac12 M^2-\frac13 g M^3$.

To analyze this path integral, it is most convenient 
to use the matrix analogue of polar coordinates.
That is, let
\be
  M(x)=\Omega(x)\Lambda(x)\Omega^{-1}(x)
\label{polarmat}
\ee
where $\Omega\in U(N)$ and $\Lambda={\rm diag}(\lam_1,\lam_2,...,\lam_N)$.
The integration measure $\DD M$ becomes in these variables
\be
  \DD M = \DD\Omega\, \DD\Lambda\, \Delta^2(\Lambda)
	\quad,\qquad \Delta(\Lambda) = \prod_{i<j}(\lambda_i-\lambda_j)
\label{vdm}
\ee
where $\DD\Omega$ is the $U(N)$ group (Haar) measure.

A useful intuition to keep in mind is the analogous transformation
from Cartesian to spherical coordinates for integration
over the vector space $\IR^n$.  One uses the rotational invariance
of the measure to write $d^n x=d\Omega_{n-1} dr\, r^{n-1}$,
with $\Omega_{n-1}$ the space of angles which parametrize
an orbit under the rotational group $O(n)$; $r$ parametrizes which
orbit we have, and $r^{n-1}$ is the size of the orbit.
The orbits degenerate at the origin $r=0$, due to its
invariance under $O(n)$, and this degeneration is responsible
for the vanishing of the Jacobian factor $r^{n-1}$
on this degenerate orbit.
Similarly, in the integration over matrices $\DD M$ is the
Cartesian measure on the matrix elements of $M$.  The invariance
of this measure under under unitary conjugation of $M$
allows us to pass to an integration over $U(N)$ orbits,
parametrized by the diagonal matrix of eigenvalues $\Lambda$.
The ({\it Vandermonde}) Jacobian factor $\Delta^2(\Lambda)$
characterizes the size of an orbit; the orbits degenerate
whenever a pair of eigenvalues coincide, since the action
of $SU(2)\subset U(N)$ (that rotates these eigenvalues
into one another) degenerates at such points.
The overall power of the Vandermonde determinant is determined
by scaling (just as the power $r^{n-1}$ is fixed for
the vector measure).

In these variables, the Hamiltonian 
for the matrix quantum mechanics \pref{mqm} is
\be
  H = \sum_i\Bigl[-\frac\beta2\frac1{\Delta^2}\frac\partial{\partial\lam_i}
		\Delta^2\frac\partial{\partial\lam_i} + \UU(M)\Bigr]
	+\frac1{2\beta}\sum_{i<j}
		\frac{\hat\Pi_{ij}\hat\Pi_{ji}}{(\lam_i-\lam_j)^2}
\label{matham}
\ee
where $\hat\Pi_{ij}$ is the left-invariant momentum on $U(N)$,
and the ordering has been chosen so that the operator
is Hermitian with respect to the measure \pref{vdm}.
The last term is the analogue of the angular momentum barrier
in the Laplacian on $\IR^n$ in spherical coordinates.
Note that the kinetic operator for the eigenvalues can be rewritten
\be
  \sum_i \frac1{\Delta^2}\frac\partial{\partial\lam_i}
        \Delta^2\frac\partial{\partial\lam_i}
  = \sum_i \frac1{\Delta}\frac{\partial^2}{\partial\lam_i^{\;2}} \Delta\ .
\label{KEident}
\ee

Wavefunctions for the $U(N)$ angular degrees of freedom
will transform in representations of $U(N)$.  The simplest
possiblity is to choose the trivial representation,
$\Psi_{U(N)}(\Omega)=1$.  In this $U(N)$ singlet sector,
we can write the wavefunction as
\be
  \Psi(\Omega,\Lambda) = \Psi_{\rm eval}(\Lambda)
	= \Delta^{-1}(\Lambda)\tilde\Psi(\Lambda)
\label{singletfn}
\ee
and the Schr\"odinger equation becomes
\cite{Brezin:1977sv}
\be
  H\Psi_{\rm eval}(\Lambda)=\Delta^{-1}(\Lambda)\sum_i\Bigl[
	-\frac\beta2\frac{\partial^2}{\partial\lam_i^{\;2}}+\UU(\lam_i)\Bigr]
	\tilde\Psi(\Lambda)\ ,
\label{schreq}
\ee
\ie\ the eigenvalues are {\it decoupled} particles moving in
the potential $\UU(\lambda)$.
The wavefunction $\Psi_{\rm eval}$ is symmetric under
permutation of the eigenvalues in the $U(N)$ singlet sector
(these permutations are just the Weyl group action of $U(N)$);
consequently $\tilde\Psi$ is totally antisymmetric under
eigenvalue permutations --
{\it the eigenvalues behave effectively as free fermions}.

\subsection{\label{nonsinglet} An aside on non-singlets}

What about non-singlet excitations?
Gross and Klebanov \cite{Gross:1990md,Klebanov:1991qa,Boulatov:1991xz}
estimated the energy cost of non-singlet excitations and found
it to be of order $O(-\log\epsilon)$, where $\epsilon\to 0$
characterizes the continuum limit.  
Hence, angular excitations decouple energetically
in the continuum limit.
Alternatively, one can gauge the $U(N)$, replacing
$\partial_x M$ by the covariant derivative
$D_xM=\partial_xM+[A,M]$; the Gauss law of the gauge theory
then projects onto $U(N)$ singlets.

The physical significance of non-singlet excitations
is exhibited if we consider the theory in periodic Euclidean time
$x\in\IS^1$, $x\sim x+2\pi R$, appropriate to the computation
of the thermal partition function.
In the matrix path integral, we must allow twisted boundary
conditions for $M$
\cite{Gross:1990md,Klebanov:1991qa,Boulatov:1991xz}:
\be
  M(x+2\pi R)=\Omega M(x)\Omega^{-1}\quad,\qquad \Omega\in U(N)\quad .
\label{twistbc}
\ee
The matrix propagator is modified to 
\be
  \vev{M_i^{~k}(x)M_j^{~l}(x')} = \sum_{m=-\infty}^\infty
	e^{-|x-x'+2\pi Rm|} \; (\Omega^m)_{i}^{~l}\,(\Omega^{-m})_j^{~k}\ .
\label{twistprop}
\ee

\begin{figure}[ht]
\begin{center}
\[
\mbox{\begin{picture}(110,110)(0,0)
\includegraphics[scale=.5]{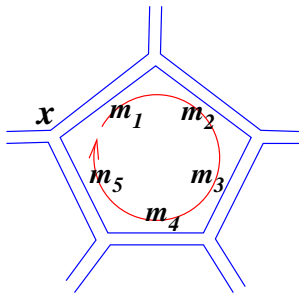}
\end{picture}}
\]
\caption{\it 
The product over twisted propagators around the face of a fatgraph
allows monodromy for $x$, corresponding to a vortex insertion.
}
\end{center}
\end{figure}

Consider a fixed set of $\{m_i\}$ and a fixed fatgraph.
Following the propagators along the index line 
that bounds the face of a planar graph,
figure \vortex, we see that the coordinate of a fatgraph
vertex along the boundary shifts by
\be
  x\longrightarrow x+2\pi\Bigl(\sum_i m_i\Bigr) R\ ;
\label{xmonod}
\ee
thus the sum over $\{m_i\}$ is a sum over vortex insertions
on the faces of the graph (the vertices of the dual tesselation).
The sum over twisted boundary conditions introduces vortices
into the partition sum for the scalar matter field $X$.
We can now understand the suppression of non-singlet
wavefunctions as a reflection of the suppression of vortices
in the 2d QFT of a periodic scalar below the Kosterlitz-Thouless transition.

\subsection{The continuum limit}

We are finally ready to discuss the continuum limit of
the sum over surfaces.  Recall that we wish to take $N\to\infty$,
with the potential tuned to the vicinity of a phase transition --
a nonanalytic point in the free energy as a function of the
couplings in the potential $\UU(M)$.  We now know that
the dynamics is effectively that of free fermionic
matrix eigenvalues, moving in the potential $\UU(\lam)$.
Consider $\UU(\lam)=-\frac12\lam^2-g\lam^3$,
figure \cubicpotl a.

\begin{figure}[ht]
\begin{center}
\[
\mbox{\begin{picture}(330,120)(0,0)
\includegraphics[scale=.5]{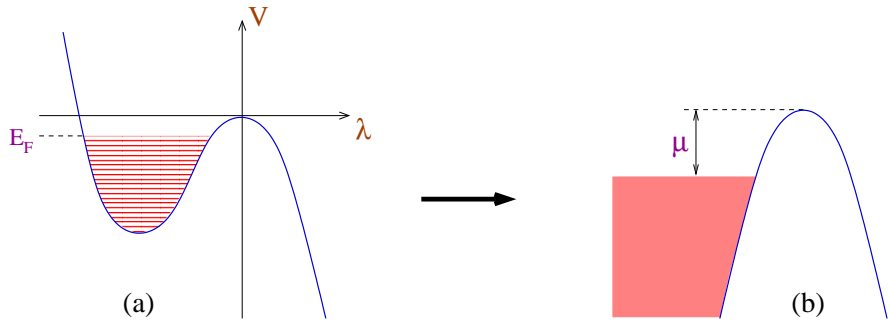}
\end{picture}}
\]
\caption{\it 
(a) Cubic eigenvalue potential.  For small $g$, there are many
metastable levels.
(b) The scaling limit focusses on the vicinity of the local maximum
of the potential.
}
\end{center}
\end{figure}

There are many metastable levels in the well 
on the left of the local maximum of the potential.
The coupling $g$ can be tuned so that there are
more than $N$ such metastable single-particle states.
As $N$ is sent to infinity, one can adjust $g\to 0$ 
so that there are always $N$ levels in the well.
The metastable Fermi energy $E_F$ will be a function of $g$ and $N$.
Consider an initial state where these states are populated
up to some Fermi energy $E_F$ below the top of the barrier,
and send $g\to 0$, $N\to\infty$, such that $E_F\to 0^-$.
In other words, the phase transition we seek is
the point where eigenvalues are about to spill over
the top of the potential barrier out of the well on the left.
The resulting situation is depicted in figure \cubicpotl b,
where we have focussed in on the quadratic maximum of
the potential via the rescaling $\hat\lambda=\lambda/\sqrt N$,
so that $\UU(\hat\lam)\sim -\frac12\hat\lam^2$.
We hold $\mu = -N E_F$ fixed in the limit.
The result is quantum mechanics of free fermions in an inverted harmonic
oscillator potential, with Fermi level $-\mu<0$.
To avoid notational clutter, we will drop the hat on the rescaled
eigenvalue, continuing to use $\lam$ as the eigenvalue coordinate
even though it has been rescaled by a factor of $\sqrt N$
from its original definition.

A useful perspective on the phase transition comes from
consideration of the classical limit of the ensemble of
eigenvalue fermions.
The leading semiclassical approximation to the degenerate
Fermi fluid of eigenvalues describes it
as an incompressible fluid in phase space
\cite{Das:1990ka,Polchinski:1991uq}.
Each eigenvalue fermion occupies a cell of volume $2\pi\hbar$
in phase space, with one fermion per cell; 
the classical limit is a continuous fluid,
which is incompressible due to Pauli exclusion.
The metastable ground state, which becomes stable
in this limit, has the fluid filling the interior
of the energy surface in phase space of energy $E_F$;
see figure \phasespace.

The universal part of the free energy comes from the
endpoint of the eigenvalue distribution near $\lam\sim 0$.
The limit $E_F\to 0^-$ leads to a change in this universal
component, due to the singular endpoint behavior 
$\rho(\lam)\sim \sqrt{\lam^2-E_F}$
of the eigenvalue density in this limit.

\begin{figure}[ht]
\begin{center}
\[
\mbox{\begin{picture}(400,160)(0,0)
\includegraphics[scale=.5]{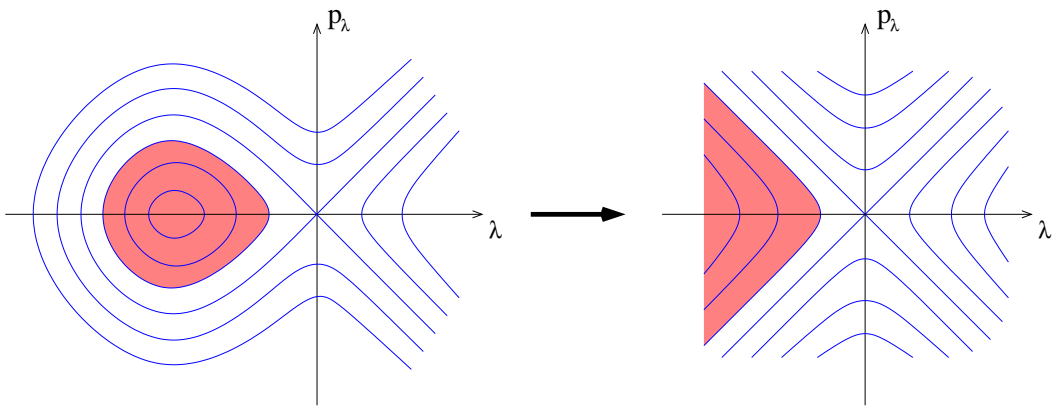}
\end{picture}}
\]
\caption{\it 
Phase space portrait of the classical limit of
the free fermion ground state.  The contours
are orbits of fixed energy; the shaded region
depicts the filled Fermi sea.
}
\end{center}
\end{figure}

One should worry that the theory we have described
is not well-defined, due to the fact that there is
a finite rate of tunnelling of eigenvalues out of the metastable well.
Single-particle wavefunctions in the inverted harmonic potential
are parabolic cylinder functions
\be
  \psi_\omega(\lam)=c_\omega D_{-\frac12+i\omega}\bigl((1+i)\lam\bigr)
	~~{\buildrel{\lam\to\infty}\over\sim}~~
	\coeff{1}{\sqrt{\pi\lam}}\, e^{-i\lam^2/2+i\omega\log|\lam|}\ .
\label{parcyl}
\ee
If we consider an incoming wave from the left with these
asymptotics, with energy $E=-\omega<0$, a WKB estimate of the
tunnelling amplitude gives $T(\omega)\sim e^{-\pi\omega}$.
Perturbation theory is an asymptotic expansion in 
$1/N\propto 1/\mu$ (from KPZ scaling), and since all
filled levels have $\omega>\mu$, tunnelling
effects behave as $e^{-cN}$ for some constant $c$
and can be ignored if one is only interested in
the genus expansion.  
The genus expansion is the asymptotic expansion around
$\mu\to\infty$, where tunnelling is strictly forbidden.%
\footnote{This is especially clear in the description 
of the classical limit as a classical, 
incompressible fluid in phase space.  The classical fluid
cannot escape the potential well via tunnelling.}
The worldsheet formalism is defined through the genus expansion;
effects such as tunnelling are invisible at fixed genus.%
\footnote{The existence of tunnelling phenomena
is reflected in the genus expansion as the rate of 
divergence of the contribution of large orders
in the expansion \cite{Zinn-Justin:1980uk};
for a discussion in the present context, see \cite{Shenker:1990uf}.}
Nonperturbatively (at finite $\mu$), the theory does not exist;
yet we can make an asymptotic expansion around the
metastable configuration of the matrix quantum mechanics, 
and compare the terms to the results of the worldsheet
path integral.  We will return to this point in
section \ref{fermstr}, where the analogous 
(and nonperturbatively stable) matrix model for the
fermionic string is briefly discussed.

The claim is that the continuum limit of the matrix path integral
just defined (valid at least in the asymptotic expansion
in $1/\mu$) is in the {\it same universality class}
as the $D=2$ string theory defined via the worldsheet
path integral for Liouville theory coupled to $c_{\rm matter}=1$
(and Faddeev-Popov ghosts).


\section{An overview of observables}

Now that we have defined the model of interest,
in both the continuum worldsheet and matrix formulations,
the next issue concerns the observables of the theory --
what physical questions can we ask?  In this section
we discuss three examples of observables:
(i) macroscopic loop operators, which put holes in the
string worldsheet; (ii) asymptotic scattering states,
the components of the S-matrix;
and (iii) conserved charges, which are present in abundance
in any free theory (\eg\ the energies of the particles
are separately conserved).

\subsection{Loops}

Consider the matrix operator
\bbb
  W(\lz,x) &=&  -\frac1N\,\tr[\log(\lz-M(x))]	
\nonumber\\
  &=& +\frac1N \sum_{l=1}^\infty \frac1l\,\tr\bigl[\bigl(M(x)/\lz\bigr)^l\bigr] 
	- \log \lz\ .
\label{matloop}
\eee
From the matrix point of view, $\exp[W(\lz,x)]=\det[\lz-M(x)]$
is the characteristic polynomial of $M(x)$, and thus a natural
collective observable of the eigenvalues.
Note that $\lz$ parametrizes the eigenvalue coordinate.
As a collective observable of the matrix, this operator
is rather natural -- its exponential is the characteristic
polynomial of the matrix $M(x)$, and hence encodes
the information contained in the distribution of
matrix eigenvalues.%
\footnote{It is amusing to note that the same operator
appears in the correspondence between $AdS_5\times\IS^5$
and maximally supersymmetric gauge theory, as the operator
that creates the gauge theory representation of
so-called ``giant gravitons'' in $AdS$
\cite{Balasubramanian:2001nh}.}
On a discretized surface, $\frac1l\tr[M^l(x)]$
is the operator that punches a hole in a surface of lattice
length $l$; see figure \macroloop.%
\footnote{The coefficient $1/l$ is a symmetry factor 
-- cyclic rotations of the legs of the vertex $\tr[M^l]$
yield the same fatgraph.}
All edges bordering the hole are pierced by a propagator
which leads to the point in time $x$ in target space,
and the other end of each propagator also goes to the point
$x$ in the continuum limit $\beta\to 0$.
Thus the continuum theory has a Dirichlet 
condition for $x$ along the boundary.

\begin{figure}[ht]
\begin{center}
\[
\mbox{\begin{picture}(150,150)(0,0)
\includegraphics[scale=.4]{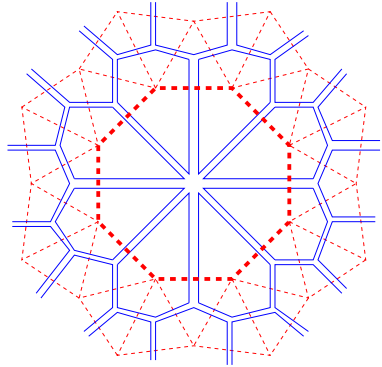}
\end{picture}}
\]
\caption{\it 
The operator $\tr[M^l(x)]/l$ inserts a boundary
of lattice length $l$ into the fatgraph ($l=8$ is depicted).
}
\end{center}
\end{figure}

It is useful to rewrite the loop operator $W(\lz,x)$ as follows:
\bbb
  W(\lz,x) &=& -\lim_{\epsilon\to 0}\int_\epsilon^\infty
	\frac{d\ell}{\ell}\, \frac1N\,\tr\,
	\exp\bigl[-\ell\bigl(\lz-M(x)\bigr)\bigr] + \log\epsilon
\nonumber\\
	&=& -\int\frac{d\ell}{\ell}\int d\lam\,
		e^{-\ell(\lz-\lam)}\hat\rho(x,\lam)+\log\epsilon
\label{Wrep}\\
	&=& -\int \frac{d\ell}{\ell}\, e^{-\ell \lz}\,
		\widetilde W(\ell,x)+\log\epsilon\quad ,
\nonumber
\eee
where in the first line we have simply 
introduced an integral representation for the logarithm,
while in the second we have rewritten the trace
over a function $f(M)$ of the matrix
as an integral over the eigenvalue coordinate $\lam$
of $f(\lam)$ times the eigenvalue density operator $\hat\rho(x,\lam)$.
This defines the operator in the third line as
\be
  \widetilde W(\ell,x) = \int d\lam\, e^{\ell\lam} \hat\rho(x,\lam)\ ,
\label{wtil}
\ee
the Laplace transform of the eigenvalue density operator
(recall that classically, the support of $\rho$
is along $(\lambda\in(-\infty,-\sqrt{2\mu})$).
The density operator is a bilinear of the fermion field operator
\bbb
  \hat\rho(x,\lam) &=& \hat\psi^\dagger\hat\psi(x,\lam)
\nonumber\\
  \hat\psi(x,\lam) &=& \int d\nu\, b_\nu\psi_\nu(\lam) e^{-i\nu x}
\label{fermop}
\eee
and its conjugate $\hat\psi^\dagger$ containing $b_\nu^\dagger$,
with the anticommutation relation of mode operators
\be
  \{b_\nu^\dagger,b_{\nu'}^{~}\}=\delta(\nu-\nu')\ .
\label{acr}
\ee
The mode wavefunctions are given in \pref{parcyl}.
The operator $\widetilde W$ is often called the
{\it macroscopic loop operator}.

In the continuum formalism, we should consider the path integral
on surfaces with boundary.  The boundary condition 
on $X$ will be Dirichlet, as discussed above.
For the Liouville field $\phi$, we use free (Neumann)
boundary conditions, but with a boundary interaction
\be
  \SS_L=\frac{1}{4\pi}\int \sqrt g[(\nabla_{\! g}\phi)^2+QR^{\sst(2)}
	+\mu\,e^{2b\phi}]
	+\sum_i\oint_{B_i}\mu_{\! B}^{(i)} e^{b\phi}\ .
\label{lioubdy}
\ee
Here, $\oint_{B_i}e^{b\phi}=\ell_{\rm \sst bdy}^{(i)}$
is the proper length of the $i^{\rm th}$ boundary
as measured in the dynamical metric; hence,
$\mu_{\! B}^{(i)}$ is the {\it boundary cosmological constant}
on that boundary component.  The path integral 
over the dynamical metric sums over boundary lengths
with the weight $e^{-\SS_{\!L}}$, and therefore
produces an integral transform with respect to the lengths
of all boundaries.  This transform has the same
structure as the last line of \pref{Wrep}.
Let us truncate to zero modes along each boundary component, 
$\ell_{\rm \sst bdy}^{\sst (i)}=e^{b\phi_0^{(i)}}$.
The path integral measure includes
$\int\! d\phi_0^{(i)}=\int\!{d\ell_{\rm \sst bdy}^{\sst (i)}}%
/{\ell_{\rm \sst bdy}^{\sst (i)}}$,
and the weight $e^{-\SS_{\!L}}$ includes
$e^{-\mu_{\! B}^{(i)}\ell_{\rm \sst bdy}^{\sst (i)}}
\PP(\ell_{\rm \sst bdy}^{\sst (i)})$,
where $\PP(\ell_{\rm \sst bdy}^{\sst (i)})$ is the probability measure
for fixed boundary lengths.  Comparison with \pref{Wrep}
suggests we identify $\ell$ in $\widetilde W(\ell,x)$
as $\ell_{\rm \sst bdy}$; $\lz=\mub$; and $\PP(\ell)$
is the correlator of a product of loop operators $\widetilde W(\ell,x)$.

Note in particular that the eigenvalue space
of $\lambda$, which by \pref{matloop} is the same as $\lz$-space,
is related to $\ell$-space (the Liouville 
coordinate $\phi$) by an {\it integral transform}.
They are not the same!  However, it is true that 
asymptotic plane waves in $\phi$ are the same as asymptotic
plane waves in $\log\lambda$.  

\subsection{The S-matrix}

Another observable is the S-matrix.
The standard worldsheet prescription for string scattering
amplitudes is to evaluate the integrated correlation functions
of on-shell vertex operators. 
Asymptotic tachyon perturbations are produced by the operators
\be
  V_{i\omega}^{\rm\sst in,out} 
	= \alpha_\pm(\omega)\,e^{i\omega(x\mp\phi)}\,e^{Q\phi}
\label{tachvert}
\ee
(whose dimension $\Delta=\bar\Delta=1$ follows from \pref{scaledim}).
The factor $e^{Q\phi}$ is just the local effective string coupling
\pref{geff}.
The vibrational modes of the string are physical only
in directions transverse to the string's worldsheet.
Since the worldsheet occupies the only two dimensions of
spacetime which are available, there are no transversely
polarized string excitations and the only physical string
states are the tachyon modes, which have only center-of-mass
motion of the string.
Actually, this statement is only true
at generic momenta.  For special momenta, there are additional
states (in fact these momenta located at the poles
in the relative normalization of $V_{i\omega}^{\rm matrix}$
and $V_{i\omega}^{\rm continuum}$).  The effects of
these extra states are rather subtle; for details,
the reader is referred to \cite{Ginsparg:1993is}.
The perturbative series for the tachyon S-matrix is 
\be
  {\bf S}\bigl(\omega_i|\omega'_j\bigr)
	=\sum_{h=0}^\infty\int\!\prod_r dm_r\;
	\Bigl\langle\prod_i\int\! d^2\!z_i\, V_{i\omega_i}^{\sst\rm (in)}\;
	\prod_j\int\! d^2\!w_j\, V_{i\omega'_j}^{\sst\rm (out)} \Bigr\rangle\ .
\label{knamp}
\ee

Actually, the statement that the tachyon is the
only physical excitation is only true at generic momenta.  
For special momenta, there are additional
states (in fact these momenta located at the poles
in the relative normalization of $V_{i\omega}^{\rm matrix}$
and $V_{i\omega}^{\rm continuum}$, see section \ref{compare}).  
The effects of these extra states are rather subtle; for details
and further references,
the reader is referred to \cite{Ginsparg:1993is}.

In the matrix approach, the {\it in} and {\it out} modes
are ripples (density perturbations)
on the surface of the Fermi sea of the asymptotic form
\be
  \delta\hat\rho(\omega,\lam) = \hat\psi^\dagger\hat\psi(\omega,\lam)
	~~~{\buildrel \lam\to-\infty \over \sim}~~~
	\frac1{2\lam}\Bigl(\alpha_+(\omega)e^{+i\omega\log|\lam|}
		+\alpha_-(\omega)e^{-i\omega\log|\lam|}\Bigr)
\label{rhoasymp}
\ee
as we will verify in the next section.
The $\alpha_\pm(\omega)$ are right- and left-moving modes
of a free field in $x\pm\log|\lam|$,
normalized as 
\be
  \bigl[\alpha_\omega^\pm,\alpha_{\omega'}^\pm\bigr]
	=-\omega\delta(\omega+\omega')\ .
\label{ccr}
\ee
Thus, to calculate the S-matrix we should perform
a kind of LSZ reduction of the eigenvalue density correlators 
\cite{Moore:1991zv}.
Once again, as in the case of the macroscopic loop,
the primary object is the density correlator.

The phase space fluid picture of the classical theory
leads to an efficient method to compute the classical S-matrix
\cite{Polchinski:1991uq,Moore:1992gb}, and provides an appealing
picture of the classical dynamics of the tachyon field.


\subsection{\label{conscharge}Conserved charges}

Since the dynamics of the matrix model is that of free fermions,
there will be an infinite number of conserved quantities
of the motion.  For instance, the energies of each of the
fermions is separately conserved.  In fact, all of
the phase space functions 
\be
  q^{~}_{mn}(\lam,p) = (\lam+p)^{r-1}(\lam-p)^{s-1}\, e^{-(r-s)x}
\label{conschg}
\ee
($p$ is the conjugate momentum to $\lam$)
are time independent for motion of a particle in the inverted
oscillator potential, generated by $H=\frac12(p^2-\lam^2)$,
ignoring operator ordering issues.  
These charges generate canonical transformations,
and can be regarded as generators of the algebra of area-preserving
polynomial vector fields on phase space
(see \cite{Ginsparg:1993is} and references therein).
Note that the time-independent operators with $m=n$
are simply powers of the energy,
$q_{mm}=(-H)^{m-1}$.
Formally, the operator
\be
  \hat q^{~}_{mn} = \int\! d\lam\, \hat\psi^\dagger (\lam)
	q^{~}_{mn}(\lam,-i\partial_\lam)\hat\psi(\lam)
\label{fieldchg}
\ee
implements the corresponding transformation on the fermion
field theory, ignoring questions of convergence.
For $m=n$ we can be more precise: Energy should be measured
relative to the Fermi energy,
\be
  \hat q^{~}_{mm} = \int_{-\mu}^\infty\! d\nu\, 
		(-\mu-\nu)^{m-1} b_\nu^\dagger b_\nu^{~}
	-\int_{-\infty}^{-\mu} \! d\nu\,
		(-\mu-\nu)^{m-1} b_\nu^{~}b_\nu^\dagger\ ;
\label{qmmhat}
\ee
this expression is finite for finite energy excitations away
from the vacuum state with Fermi energy $-\mu$.

The operators realizing these conserved charges 
in the worldsheet formalism were exhibited in \cite{Witten:1991zd}
(for recent work, see 
\cite{Sen:2004zm,Sen:2004yv}).
The charges $q_{12}$ and $q_{21}$ generate the full
algebra of conserved charges, so it is sufficient
to write expressions for them.  They
are realized on the worldsheet as operators
$\OO_{12}$ and $\OO_{21}$
\bbb
\CO_{12}&=&(\cc\bb+\partial\phi-\partial x)
(\bar \cc \bar \bb+\bar\partial\phi-\bar\partial x)e^{-x-\phi}
\nonumber\\
\CO_{21}&=&(\cc\bb+\partial\phi+\partial x)
(\bar \cc \bar \bb+\bar\partial\phi+\bar\partial x)e^{+x-\phi}\ .
\label{gringops}
\eee
Here $\bb(z)$, $\cc(z)$ are the Faddeev-Popov ghosts
for the local gauge choice $g_{ab}=\delta_{ab}$,
\cf\ \cite{Green:1987sp,Polchinski:1998rq}.
These operators have scale dimension $\Delta=\bar\Delta = 0$,
and can be placed anywhere (unintegrated) 
on the two-dimensional worldsheet --
moving them around changes correlators by gauge artifacts
which decouple from physical quantities.
The relation between matrix and continuum expressions for the
conserved charges was worked out recently in 
\cite{Sen:2004zm,Sen:2004yv}.


\section{\label{diskcalc}Sample calculation: the disk one-point function}

An illustrative example which will allow us to compare
these two rather different formulations of 2D string theory
(and thereby check whether they are in fact equivalent)
is the mixed correlator of one in/out state and one
macroscopic loop.
This correlator computes the process whereby an incoming
tachyon is absorbed by the loop operator
(or an outgoing one is created by the loop).

\subsection{Matrix calculation}

On the matrix side, we must evaluate the density-density
correlator
\be
  \bra{{\rm vac}} \hat\rho(\lam_1,x_1)\,\hat\rho(\lam_2,x_2) \ket{{\rm vac}}
\label{densdens}
\ee
and Laplace transform with respect to $\lam_1$ to get
the macroscopic loop, while performing LSZ reduction
in $\lam_2$.  The evaluation of \pref{densdens}
proceeds via substitution of \pref{fermop} and 
use of \pref{acr} as well as the vacuum property
\bbb
  b_\nu\ket{{\rm vac}}=0\quad,\qquad \nu>\mu
\nonumber\\
  b_\nu^\dagger\ket{{\rm vac}}=0\quad,\qquad \nu<\mu
\label{vacdef}
\eee
(note that we have not performed the usual redefinition
of creation/annihilation operators below the Fermi surface).
The result is \cite{Moore:1991sf}
\be
  \vev{\hat\rho(1)\hat\rho(2)} =
	\int_\mu^\infty d\nu \, e^{-i\nu(x_2-x_1)}
	\psi_\nu^\dagger(\lam_1)\psi_\nu^{~}(\lam_2)
	\int_{-\infty}^\mu d\nu'\, e^{i\nu'(x_2-x_1)}
	\psi_{\nu'}^{~}(\lam_1)\psi_{\nu'}^\dagger(\lam_2)\ .
\label{ddresult}
\ee
The parabolic cylinder wavefunctions have the asymptotics
(for $Y=\sqrt{\lam^2-2\nu}\gg 1$, $\nu\gg 1$)
\be
  \psi_\nu(\lam)\sim\Bigl[\frac{1}{\pi Y}\Bigr]^{1/2}\,
	\sin\Bigl(\coeff12 \lam Y+\nu\tau(\nu,\lam)-\coeff\pi4\Bigr)
\label{parcylas}
\ee
where
\be
  \tau(\nu,\lam) = -\int_{-2\sqrt\nu}^{-\lam}
	\frac{d\lam'}{\sqrt{\lam^{\prime\,2}-2\nu}}
	=\log\Bigl(\frac{-\lam+\sqrt{\lam^2-2\nu}}{\sqrt{2\nu}}\Bigr)
\label{taudef}
\ee
is the WKB time-of-flight of the semiclassical fermion trajectory,
as measured from the turning point of its motion.

At this point, we will make some approximations.
We wish to compare the matrix and worldsheet field theory
computations.  However, the latter is only well-behaved
in a low-energy regime, as we saw in section \ref{ddimstr}.
Therefore we will approximate the energies in \pref{ddresult}
as $\nu\sim\mu+\delta$, $\nu'\sim\mu-\delta'$,
with $\delta,\delta'\ll\mu$, so that the density perturbation
is very near the Fermi surface.  In addition, 
substituting the parabolic cylinder wavefunction
asymptotics \pref{parcylas}
in \pref{ddresult}, we drop all rapidly oscillating terms
going like $\exp[\pm \frac i2\lam^2]$; these
terms should wash out of the calculation when we take
$\lam_2\to\infty$ to perform the LSZ reduction.

With these approximations, one finds
\be
  \psi_\nu(\lam_2)^{~}\psi^\dagger_{\nu'}(\lam_2)
	~~~{\buildrel \lam_2\to -\infty \over\sim}~~~
	\frac{1}{4\pi\lam_2}\Bigl[
		\Bigl(\!\sqrt{\coeff{2}{\mu}}\,|\lam_2|\Bigr)^{i(\nu-\nu')}
		+\Bigl(\!\sqrt{\coeff{2}{\mu}}\,|\lam_2|\Bigr)^{-i(\nu-\nu')}
	\Bigr]+O\Bigl(\frac{\omega^2}{\mu}\Bigr)
\label{lszlamtwo}
\ee
(recall $\omega=\nu-\nu'$).  We wish to identify this with the
in/out wave \pref{rhoasymp}.  Recall that initially the
wavefunctions were multiplied by mode operators $b_{\nu'}^\dagger$,
$b_\nu^{~}$; there is also a sum over energies.
Comparing, we see that
\be
  \alpha_\omega = \int_0^\omega d\vareps\, b_{\omega-\vareps}^\dagger
	b_\vareps\times\frac1{2\pi}\Bigl(\frac\mu 2\Bigr)^{-i\omega/2}
\label{bosonize}
\ee
which is (up to an overall phase, which we can absorb
in the definition of the operators) just the standard
{\it bosonization formula} for 2D fermions.%
\footnote{The fermions are asymptotically relativistic
in $t\pm\log|\lam|$ as $\lam\to-\infty$.}

As for the other part of the expression, the wavefunctions at $\lam_1$,
we make the same set of approximations, except that we use
the full expression \pref{parcylas} rather than its $\lam\to\infty$
limit.  One finds
\be
  \psi^\dagger_\nu(\lam_1)\psi^{~}_{\nu'}(\lam_1) \sim
	\coeff{1}{4\pi\sqrt{\lam_1^2-2\mu}}\Bigr[
	\Bigl(\coeff{-\lam_1+\sqrt{\lam_1^2-2\mu}}{\sqrt{2\mu}}
		\Bigr)^{i(\nu-\nu')}  
	+\Bigl(\coeff{-\lam_1+\sqrt{\lam_1^2-2\mu}}{\sqrt{2\mu}}
		\Bigr)^{-i(\nu-\nu')} 
	\Bigr]+O\Bigl(\frac{\omega^2}{\mu}\Bigr)\ .
\label{lamonered}
\ee

Note that the terms of order $\omega^2/\mu$ that have been dropped
are exactly of the form to be contributions of higher
topologies of worldsheet.  As we saw in the scattering
of waves bouncing off the exponential Liouville wall
in section \ref{ddimstr}, the effective string coupling
\pref{geffmax} is $g_{\rm eff}\sim \omega^2/\mu$.

Fixing the sum of the energies $\nu-\nu'=\omega$
(\eg\ by Fourier transformation in $x$), the remaining
energy integral is trivial and gives a factor of $\omega$.
The macroscopic loop is finally obtained by Laplace transform
with respect to $\lam_1$; the answer is a Bessel function:%
\footnote{The wavefunctions $\psi_\nu(\lam)$ are exponentially
damped under the barrier for $\nu\sim -\mu$, and so we
may approximate the range of integration as 
$\lam\in(-\infty,-\sqrt{2\mu})$.}
\be
  \int_1^{\infty}\Bigl[\Bigl(\sqrt{t^2-1}+t\Bigr)^{i\omega}
	+\Bigl(\sqrt{t^2-1}+t\Bigr)^{-i\omega}\Bigr]
	\, e^{-ut}\,\frac{dt}{\sqrt{t^2-1}}
	= 2\,K_{i\omega}(u)
\label{besselrep}
\ee
so that 
\be
  \widetilde W_{i\omega}(\ell)\equiv 
  \,_{\rm out}^{~}\bra{{\rm vac}}\,
	\widetilde W(\ell,x)\,\ket{\omega}^{~}_{\rm in}
	= 2\,\frac{\omega}{2\pi}\,K_{i\omega}(\sqrt{2\mu}\,\ell)\ .
\label{wtilamp}
\ee
The transformation \pref{Wrep} to $z$-space yields
\bbb
  W_{i\omega}(\lz)&\equiv& \,_{\rm out}^{~}\bra{{\rm vac}}\,
	W(\lz,x)\,\ket{\omega}^{~}_{\rm in}
	= \int_0^\infty\frac{d\ell}\ell\,
		e^{-\ell\sqrt{2\mu}\,\ch\pi s}\,\widetilde W_{i\omega}(\ell)
\nonumber\\
	&=& 2\,\frac{\omega}{2\pi}\Gamma(i\omega)\Gamma(-i\omega)\,
		\cos(\pi s\omega)
\label{wamp}
\eee
where we have parametrized $\lz=\sqrt{2\mu}\,\ch(\pi s)$.

The amplitude just calculated actually reveals quite a bit
about the theory.  
We have learned that the corrections to the leading-order
expressions \pref{wtilamp}, \pref{wamp} are of order
$\omega^2/\mu$, in agreement with the estimated higher
order corrections in Liouville theory.  It is a straightforward
(if tedious) exercise to retain higher orders in the 
expansion, and thereby compute the corrections
to the amplitude coming from surfaces with handles.

Another feature of Liouville theory we see appearing is its
quantum wavefunction
\cite{Moore:1991ir,Moore:1991ag}.  
In quantum theory, an operator $\OO$ creates a state $\OO\ket{0}$,
whose overlap with the position eigenstate $\ket{x}$
is the wavefunction $\psi_\OO^{\,}(x)=\bra{x}\OO\ket{0}$.
Similarly, we wish to interpret the 
state created by the macroscopic loop
$\widetilde W(\ell,x)\ket{\rm vac}$ as the position eigenstate 
in the space of $(\ell,x)$, whose overlap with the
state $V_{i\omega}\ket{\rm vac}$ is the wavefunction corresponding
to the operator $V_{i\omega}$.  This wavefunction is sometimes
called the {\it Wheeler-de Witt} wavefunction.

In the continuum formulation
the correlation function \pref{wamp} involves one
macroscopic loop of boundary cosmological constant $\mub$,
and one tachyon perturbation $V_{i\omega}$, as depicted
in figure \wavefn.

\begin{figure}[ht]
\begin{center}
\[
\mbox{\begin{picture}(150,123)(0,0)
\includegraphics[scale=.6]{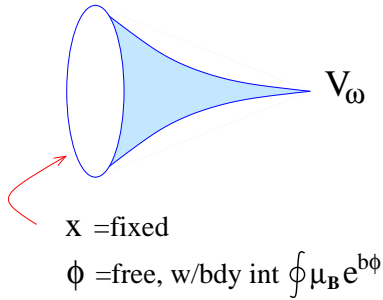}
\end{picture}}
\]
\caption{\it 
The disk one-point function of a tachyon perturbation
is the leading-order contribution to the process whereby
an incoming tachyon is absorbed by a macroscopic loop operator.
}
\end{center}
\end{figure}

Indeed, if we butcher the theory by truncating to the
spatial zero modes 
$\phi_0(\sigma_0)=\frac1{2\pi}\int\! d\sigma_{\!1} \phi(\sigma_0,\sigma_1)$ 
on a worldsheet of cylindrical topology,%
\footnote{More precisely, a semi-infinite cylinder.
The finite boundary of the cylinder is the macroscopic loop,
while the end of the cylinder at infinity
(conformal to a puncture on the worldsheet) is where
the vertex operator $V_{i\omega}$ creates the incoming asymptotic state.}
we arrive at Liouville quantum mechanics,
whose Schr\"odinger equation reads
\be
  \Bigl[-\frac{\partial^2}{\partial\phi_0^{\;2}}
	+ 2\pi\mu\,e^{2b\phi_0} \; -\;\omega^2\Bigr]
	\;\psi_\omega(\phi_0) = 0\ .
\label{liouqm}
\ee
The resulting wavefunctions
\be
  \psi_\omega(\phi_0) = \frac{2\,(\mu/2)^{-i\omega/2b}}{\Gamma(-i\omega/b)}
	\; K_{i\omega/b}\Bigl(\!{\sqrt{2\mu}}\,e^{b\phi_0}\Bigr)
\label{LQMwavefn}
\ee
are, up to normalization, identical to $\widetilde W(\ell=e^{b\phi_0})$.

\subsection{\label{wdw}Continuum calculation}

There is actually more to be learned from the exact
evaluation of this disk one-point correlator
in the full Liouville plus matter CFT,
as opposed to its quantum-mechanical zero mode truncation.
In particular, one finds the precise relation between
equivalent observables of the two formalisms.
The non-trivial part is the calculation of the
Liouville component, which rests on
a conformal bootstrap for Liouville correlators on surfaces
with boundary developed in
\cite{Fateev:2000ik,Teschner:2000md,Zamolodchikov:2001ah},
building on earlier work 
(reviewed in \cite{Teschner:2001rv,Nakayama:2004vk})
on closed surfaces.
We will only sketch the construction; the reader interested
in more details should consult these references
(and the references in these references).

The basic observation is the identity
\be
  \partial^{\;2}_z \V_{-b/2}(z) = b^2\,T_{zz}\,\V_{-b/2}(z)
\label{wardident}
\ee
(and similarly for $\V_{\!-1/2b}$, \ie\ $b\leftrightarrow 1/b$),
where $\V_\alpha=e^{2\alpha\phi}$ are the exponential operators
of Liouville field theory.
This identity is consistent with the semiclassical limit
$b\to 0$, $Q=b^{-1}+b\to b^{-1}$, since
\be
  \partial^{\;2}_z \,e^{-b\phi}=
	[b^2(\partial_z\phi)^2-b\partial_z^{\;2}\phi]\,e^{-b\phi}
	=b^2 T_{zz}\,e^{-b\phi}\ .
\label{classWI}
\ee
Correlation functions with extra insertions of $T_{zz}$
are given in terms of those without such insertions,
by the Ward identities of conformal symmetry.
Thus, plugging \pref{wardident} into a correlation function
leads to second order differential equations
on correlators involving $\V_{-b/2}$ (and similarly $\V_{\!-1/2b}$).
Conformal invariance also dictates the structure of the correlator
we wish to calculate,
\be
  \vev{\V_\alpha(z)}^{~}_{\mub}
	= \frac{U(\alpha)}{|z-\bar z|^{2\Delta_\alpha}}
\label{oneptform}
\ee
where $z$ is a coordinate on the upper half-plane,
see figure \oneptfn.
This is equivalent to the correlator on the
disk via the conformal transformation $z=-i\frac{w+i}{w-i}$;
taking into account that the operator $\V_\alpha$ transforms
like a tensor of weight $\Delta_\alpha$
in both $z$ and $\bar z$, one finds
\be
  \vev{\V_\alpha(z)}^{~}_{\mub,\;{\rm disk}}
	=\frac{U(\alpha)}{(1-|w|^2)^{2\Delta_\alpha}}\ .
\label{diskonept}
\ee
The nontrivial information lies in the
overall coefficient $U(\alpha)$.

\begin{figure}[ht]
\begin{center}
\[
\mbox{\begin{picture}(135,100)(0,0)
\includegraphics[scale=.5]{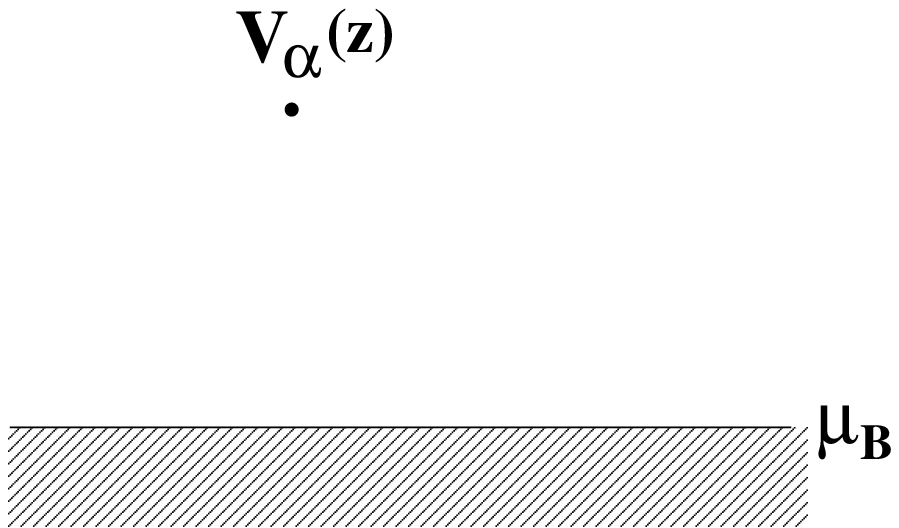}
\end{picture}}
\]
\caption{\it 
}
\end{center}
\end{figure}

In order to employ the Ward identity \pref{wardident},
we consider instead the two-point correlator
\be
  \vev{\V_\alpha(z)\, \V_{-b/2}(w)}\ .
\label{twopt}
\ee
The fact that $\V_{-b/2}$ satisfies a second order
differential equation implies that only two scaling dimensions
(up to integers) appear in its operator product expansion (OPE)
with $\V_\alpha$, schematically
\be
  \V_\alpha \V_{-b/2} \sim
	C_+(\alpha)\Bigl[\V_{\alpha-b/2}\Bigr]
	+C_-(\alpha)\Bigl[\V_{\alpha+b/2}\Bigr]\ ,
\label{twoblocks}
\ee
where the square brackets denote the operator together
with all that can be obtained from it by the action
of the conformal algebra, the so-called {\it conformal block}.
The differential equation coming from \pref{wardident},
together with \pref{twoblocks}, yields
\be
  \vev{\V_\alpha(z)\, \V_{-b/2}(w)} =
	C_+(\alpha)U(\alpha-b/2)\GG_+(\xi)
	+C_-(\alpha)U(\alpha+b/2)\GG_-(\xi)
\label{twoptsoln}
\ee
where $\GG_\pm$ are hypergeometric functions of the cross-ratio 
$\xi=\frac{(z-w)(\bar z-\bar w)}{(z-\bar w)(\bar z-w)}$.
What are the coefficients $C_\pm(\alpha)$?
For $C_+(\alpha)$ the result of the OPE satisfies
conservation of the ``charge'' of the exponential
(the Liouville zero-mode momentum $p^{~}_\phi$).
Even though this momentum is not conserved due to 
the presence of the tachyon wall, which violates
translation invariance in $\phi$,
if we nevertheless use free field theory to evaluate it
we find trivially $C_+(\alpha)=1$.
We similarly use naive perturbation theory in powers of $\mu$
to evaluate $C_-(\alpha)$, bringing down the tachyon
potential $\int\! \mu e^{2b\phi}$ in a power series expansion
and evaluating the resulting integrated correlation
functions using free field theory.  Only the first term 
in the $\mu$ expansion contributes,
and we find
\bbb
  C_-(\alpha) &=& \Bigl\langle 
	\V_\alpha(0)\V_{-b/2}(1)\V_{Q-\alpha-b/2}(\infty)
	\Bigl(-\mu\int d^2z\, \V_b(z,\bar z)\Bigr)\Bigr\rangle_{\rm FFT}
\nonumber\\
	&=& -\pi\mu\,\frac{\gamma(2b\alpha-1-b^2)}{\gamma(-b^2)
		\gamma(2b\alpha)}
\label{cminus}
\eee
where $\gamma(x)\equiv \frac{\Gamma(x)}{\Gamma(1-x)}$.%
\footnote{$\V_{\!Q-\alpha-b/2}$ is the conjugate operator to
$\V_{\!\alpha+b/2}$ in free field theory with a linear dilaton
of slope~$Q$.}

Why are we allowed to use a perturbative expansion in $\mu$
and free field theory for evaluating these quantities?
After all, the loop amplitude 
$\mu^{-i\omega/2}K_{i\omega}(\sqrt{2\mu}\,\ell)$
is certainly {\it not} polynomial in $\mu$.
Nevertheless, for special ``resonant'' amplitudes 
this procedure is justified.
Resonant amplitudes are
those for which the sum of the exponents 
$\sum \alpha_i$ of the collection of Liouville operators
$\V_{\alpha_i}$ adds up to a negative 
multiple of the exponent $2b$ of the Liouville potential.
In such cases, the 
path integral can be evaluated by perturbation theory in $\mu$.
This feature is related to the property that the integral over 
the constant mode of $\phi$ in the path integral 
is dominated by the
region $\phi\to -\infty$, where the Liouville potential
is effectively vanishing.  The use of
free field theory methods is then justified.  
The correlators that define $C_\pm$ 
satisfy this resonance condition.
Note that we are not using free field theory to evaluate the full amplitude,
but rather only to evaluate the operator product coefficients
with the special degenerate operator $\V_{-b/2}$
(and similarly $\V_{\!-1/2b}$).

We now have partial information on the correlation function.
To get a closed system of equations, we need a second
relation on \pref{twopt}.  For this purpose,
we consider the OPE of $\V_{-b/2}(w)$ with its image
across the boundary to make the identity operator,
by taking $w\to\bar w$
(in the process, we need to transform to another basis
for the hypergeometric functions $\GG_\pm$ adapted
to this particular degeneration).
In this limit, the correlator \pref{twopt} factorizes,
\be
  \vev{\V_\alpha(z)\V_{-b/2}(w)}_{\mub}
	~~~{\buildrel w\to\bar w\over \sim}~~~
	\vev{\V_{\alpha}(z)}_{\mub}\;\vev{\V_{-b/2}(w)}_{\mub}\ .
\label{fctrzn}
\ee
The first factor on the right-hand side is given
simply in terms of $U(\alpha)$, and the second factor
is yet another resonant amplitude, which we can evaluate
in free field theory by bringing down the
boundary cosmological constant interaction
from the action:
\be
  ({\rm Im}\,w)^{2\Delta_\alpha}\,\Bigl\langle
	\V_{-b/2}(w) B_Q(\infty)\Bigl(-\mub\oint\! d\xi B_b(\xi)\Bigr)
	\Bigr\rangle_{\rm FFT}
	= -2\pi\mub\,\frac{\Gamma(1-2b^2)}{\Gamma^2(-b^2)}\ .
\label{bdyresamp}
\ee
Here the integral over $\xi$ is along the boundary,
which is the real axis; $B_\alpha$ is the 
operator $e^{2\alpha\phi}$ inserted on the boundary;
and $B_{Q}$ represents the extrinsic curvature of the boundary
at infinity.

Equating the two expressions \pref{twoptsoln} and \pref{fctrzn},
and using \pref{cminus}, \pref{bdyresamp}, one arrives
at a shift relation on $U(\alpha)$
\cite{Fateev:2000ik,Teschner:2000md,Zamolodchikov:2001ah},
\be
  -\frac{2\pi\mub}{\Gamma(-b^2)}\,U(\alpha)
	= \frac{\Gamma(-b^2+2b\alpha)}{\Gamma(-1-2b^2+2b\alpha)}
		\,U(\alpha-b/2)
	-\frac{\pi\mu\Gamma(-1-b^2+2b\alpha)}{\gamma(-b^2)\Gamma(2b\alpha)}
		\,U(\alpha+b/2)\ .
\label{shiftrel}
\ee
There is a similar shift relation obtained by use of
$\V_{\!-1/2b}$.  It is convenient to write $\mub$ in
terms of a parameter $s$ via
\be
  \cosh^2(\pi bs) = \frac{\mu_{\!B}^{\;2}}{\mu}\,\sin(\pi b^2)\ ;
\label{mubparam}
\ee
then the two discrete shift relations (obtained
by use of both $\V_{-b/2}$ and $\V_{-1/2b}$) are solved by
\be
  U(\alpha) = \frac 2b\Bigl(\pi\mu\gamma(b^2)\Bigr)^{\frac{Q-2\alpha}{2b}}
	\Gamma(2b\alpha-b^2)\Gamma(\frac{2\alpha}{b}-\frac1{b^2}-1)
	\,\cosh[(2\alpha-Q)\pi s]\ .
\label{Usoln}
\ee
For the vertex operators with $\alpha=\half Q+\frac i2\omega$
appearing in the scattering amplitudes, this translates into
\be
  U(\alpha=\coeff12 Q+\coeff i2\omega) = 
	2i\omega\,\Bigl(\pi\mu\gamma(b^2)\Bigr)^{-i\omega/2b}\,
	\Gamma(ib\omega)\Gamma(i\omega/b)\,\cos(\pi s\omega)\ .
\label{Uscatt}
\ee
The shift operator relations don't fix the overall normalization
of $U(\alpha)$.  This normalization is obtained by demanding
that the residues of the poles at 
$2\alpha = Q-nb$ (\ie, $i\omega=-nb$) for $n=1,2,3,...$,
agree with the ``resonant amplitude'' integrals
for these special momenta.

Note that the full set of resonant amplitude integrals
involve bringing down powers of both
$\mu e^{2b\phi}$ and also $\tilde\mu e^{(2/b)\phi}$
from the action.  One needs to use the complete set 
in order to provide sufficient constraints to 
fully determine the Liouville correlators.
Hence {\it both} are present in the theory;
moreover, one finds for consistency that their coefficients
must be related:
\be
  \pi\tilde\mu\,\gamma(1/b^2) = [\pi\mu \,\gamma(b^2)]^{1/b^2}\ .
\label{mutilrel}
\ee
It turns out that this is more or less the relation
implied by the analytic continuation of the amplitude
for reflection off the Liouville potential
\be
  \tilde\mu/b = \mu\,b \,\RR(\omega=i(Q-2b))\ .
\label{murefl}
\ee
The reflection amplitude $\RR(\omega)$ for $V_{i\omega}\to V_{-i\omega}$
may be read off the two-point correlation function for tachyon 
vertex operators.
A similar relation holds for the boundary cosmological constant;
the boundary interaction is actually
\be
  \delta \SS_{\rm bdy} = \oint\Bigl(
	\mub e^{b\phi}+\tilde\mub e^{(1/b)\phi}\Bigr)
\label{truebdyint}
\ee
with
\be
  \cosh^2(\pi s/b) = \frac{\tilde\mu_{\!B}^{\;2}}{\tilde\mu}\,\sin(\pi/b^2)\ .
\label{mubtilrel}
\ee
Thus there is a kind of strong/weak coupling duality in 
Liouville QFT, characterized by
\be
  b\leftrightarrow 1/b
	\quad,\qquad
	\mu\leftrightarrow\tilde\mu 
	\quad,\qquad
	\mub\leftrightarrow\tilde\mub
\label{strwk}
\ee
(recall that $b\to 0$ was the weak coupling limit
of Liouville theory).
The parameter $s$ is invariant under this transformation.

\subsection{\label{compare}Comparing the results}

Finally, we are ready to compare the two approaches.
First we must assemble the Liouville disk amplitude
with the contributions of the free matter field $X$
and the Faddeev-Popov ghosts.  There is a factor
of $1/2\pi$ from gauge fixing the conformal isometries
of the punctured disk (rotations around the puncture).
The disk expectation value of the matter is
\be
  \Bigl\langle e^{i\omega X}(z)\Bigr\rangle_{\rm Dirichlet}
	= \frac{1}{|z-\bar z|^{2\Delta_\omega}}
\label{Xcorr}
\ee
(equivalently, $(1-|w|^2)^{-2\Delta_\omega}$ 
if we are working on the disk
rather than the upper half-plane),
which simply reflects the fact that the Dirichlet
boundary condition on $X$ is a delta function
(and hence its Fourier transform is one).
The factors of $|z-\bar z|$ cancel among Liouville, matter,
and ghosts (we must take $b\to 1$ in the Liouville part
since $D=2$; this involves a multiplicative renormalization
of $\mu$ and $\mub$ in order to obtain finite results).
This cancellation of coordinate dependence 
merely reflects that we have correctly
calculated a conformally invariant and therefore physical amplitude.
Thus
\be
  \vev{V_{i\omega}(z,\bar z)}_{\rm disk} = \frac{1}{2\pi}
	\, 2i\omega\,\muhat^{-i\omega/2}
	(\Gamma(i\omega))^2\,\cos(\pi s\omega)
\label{contonept}
\ee
where we have defined
\be
  \muhat = \pi\mu \gamma(b^2) 
	~~~{\buildrel b\to1\over\sim}~~~
	2\pi\mu(1-b)
\label{muhatdef}
\ee 
as the quantity to be held fixed in the $b\to 1$ limit.

Comparing to the matrix model result \pref{wamp},
we find the same result provided that we identify
\bbb
  V_{i\omega}^{\rm matrix} &=& 
	(\muhat)^{i\omega/2}\,\frac{\Gamma(-i\omega)}{\Gamma(i\omega)}
		\, V_{i\omega}^{\rm continuum}
\nonumber\\
  \coeff12\mu^{\rm matrix} &=& 
	\muhat^{\rm continuum}
\label{paramrels}\\
  \coeff12\mu_{\! B}^{\rm matrix} &=& \coeff12 \lz_{\rm matrix}
	=\hat\mu_{\!B}^{\rm continuum} \equiv 2\pi\mu_{\!B}^{\rm cont}(1-b)\ .
\nonumber
\eee
(the last relation amounts to $\mubhat=\sqrt{\muhat}\,\ch(\pi s)$).
Thus the exact evaluation of the worldsheet amplitude
allows a precise mapping between the continuum and matrix approaches.

The energy-dependent phase in the relative normalizations
of $V_{i\omega}$ results in a varying time delay
of reflection for particles of different energy.
It was shown in 
\cite{Natsuume:1994sp}
that this time delay reproduces what one would expect
based on the gravitational redshift seen by one particle
after another has been sent in.  Thus the so-called
``leg-pole factor'' $\frac{\Gamma(-i\omega)}{\Gamma(i\omega)}$
in equation \pref{paramrels}
is an important physical effect, which is added by hand
to the matrix model.  It is not yet understood if there
is a derivation of this factor from first principles
in the matrix model.

Other amplitudes that have been computed on both sides
of the correspondence and shown to agree include
\begin{itemize}
\item
The tree level S-matrix
\cite{DiFrancesco:1991ss,DiFrancesco:1991ud,Moore:1992gb},%
\footnote{The leg-pole factor \pref{paramrels}
in the relative normalization of vertex operators
was first observed here, and shown to be a property
of all the tree amplitudes.}
\item
The torus partition function
\cite{Bershadsky:1990xb,Sakai:1990ag},
\item
The disk one-point function calculated above
\cite{Moore:1991sf,%
Moore:1991ir,%
Moore:1991ag,%
Fateev:2000ik,%
Teschner:2000md},
\item
The annulus correlation function for two macroscopic loops
\cite{Moore:1991ir,Martinec:2003ka}.
\end{itemize}
One can also show that the properties of
the ground ring of conserved charges defined in section \ref{conscharge}
agree between the matrix and continuum formulations,
at leading order in $1/\mu$ \cite{Douglas:2003up}.
For instance, on the sphere one calculates using
the Liouville OPE coefficients $C_\pm$ that
\be
  \vev{\OO_{12}\OO_{21}}=\vev{\OO_{22}}=\vev{-H} =\mu\ .
\label{gringcorr}
\ee
This result is consistent with the fact that perturbative
excitations live at the Fermi surface, 
where the energy is $H=\frac12(p_{\lam}^{\,2}-\lam^2)=-\mu$.

Thus the matrix approach is reproducing the quantum
dynamics of Liouville CFT coupled to a free field.
Note that the matrix approach is much more economical
computationally, and we immediately see how to compute
the higher order corrections (just go to higher order
in $1/\mu$ in our approximations); for Liouville,
we need to work {\it much} harder -- we need to
go back to the conformal bootstrap and compute correlation
functions on the disk with handles, then integrate
over the moduli space.


\section{Worldsheet description of matrix eigenvalues}

Finally, what about the eigenvalues themselves?  They
are gauge invariant observables which are manifest in
the matrix formulation; what is their description in
the continuum formalism?  Note that this question
bears on the continuum description of nonperturbative
phenomena such as the eigenvalue tunnelling which leads
to the nonperturbative instability of the model.
Experience from string theory in higher dimensions
(\eg\ black hole microphysics) has taught us that
D-brane dynamics provides a description of strong coupling
physics.  Therefore we should examine the D-branes of 2D
string theory.  The fact that the tension
of D-branes is naively $O(1/g_s)$ means that they are
the natural light degrees of freedom in the strong coupling region.

In the worldsheet description of dynamics, a D-brane is
an object which puts boundaries on the worldsheet.  
The boundary conditions on the worldsheet fields $X^\mu$
tell us about the position of the brane
and the boundary interactions in the worldsheet action
specify the background fields localized on the brane.  
Perturbations of the boundary background fields are
(marginal) scaling operators on the boundary.
The theory thus has two sectors of strings --
{\it open strings} that couple to worldsheet boundaries (D-branes),
and {\it closed strings} that couple to the bulk
of the worldsheet.

In a sense, the macroscopic loop is a spacelike D-brane --
one with Dirichlet boundary conditions in the timelike direction $X$
and Neumann boundary conditions in the spacelike direction $\phi$.
The boundary interaction $\mub\oint e^{b\phi}$ is
a ``boundary tachyon'' that keeps $\phi_{\rm bdy}$
away from the strong coupling region $\phi\to\infty$
(at least for the appropriate sign of $\mub$).
This D-brane is however a collective observable at fixed time $x$
of the matrix model, and not a dynamical object.
A depiction of the D-brane interpretation of the calculation
of section \ref{diskcalc} is shown in figure \loopamps a.

\begin{figure}[ht]
\begin{center}
\[
\mbox{\begin{picture}(351,160)(0,0)
\includegraphics[scale=.6]{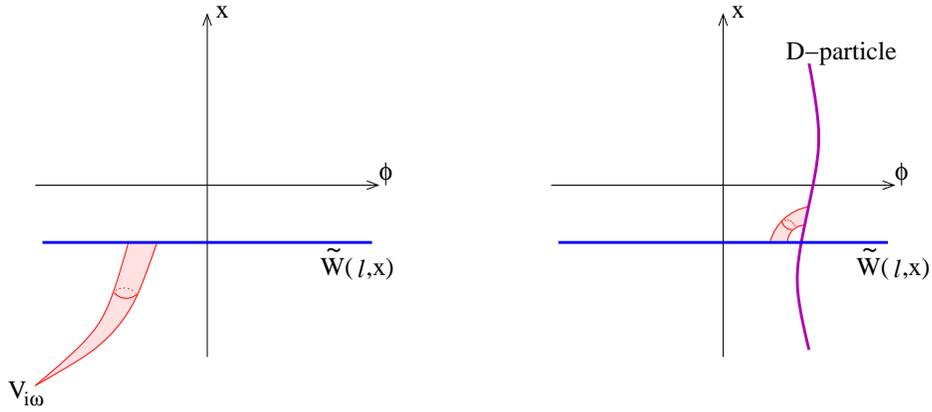}
\end{picture}}
\]
\caption{\it 
(a) A macroscopic loop is a spatial D-brane that absorbs
and emits closed strings.
(b) The loop is also a probe of the motion of D-particles.
}
\end{center}
\end{figure}

Instead, the matrix eigenvalue is localized in the spatial coordinate
$\lambda$ and hence quasi-localized in $\phi$.
Here it is important to recall that $\phi$ and $\lambda$
are related by the integral transform \pref{Wrep},
and are thus not directly identified.  Nevertheless,
localized disturbances in $\phi$ bouncing off the exponential
Liouville wall are related to localized disturbances of
the Fermi surface in $\lambda$ bouncing off the inverted
oscillator barrier, so there is a rough equivalence.

Therefore, we consider Dirichlet boundary conditions for $\phi$.
Since $\phi$ shifts under local scale transformations
($e^{2b\phi}g_{ab}$ is the dynamical metric),
the Dirichlet boundary condition
\be
  \Bigl.\phi\Bigr|_{\rm bdy} = \phi_0
\label{dbcphi}
\ee
is not conformally invariant unless $\phi=\pm\infty$.
Now $\phi=-\infty$ is the weak coupling asymptotic boundary
of $\phi$ space, and corresponds to boundaries of zero size,
which we usually think of as punctures in the worldsheet where
local vertex operators are inserted.  On the other hand,
$\phi=+\infty$ is what we want, a boundary
deep inside the Liouville wall at strong coupling.

In fact, we know a classical (constant negative curvature)
geometry with this property:
\be
  ds^2 = e^{2b\phi}dzd\bar z = 
	\frac{Q}{\pi\mu b}\frac{dzd\bar z}{(1-z\bar z)^2}\ ,
\label{poindisk}
\ee
the Poincar\'e disk (or Lobachevsky plane).  Proper distances
blow up toward the boundary: $\phi\to\infty$,
as advertised.

For this D-brane to move in time, the boundary condition in $X$
should be Neumann.  What sort of conformally invariant boundary
interaction can we have?  Since $\phi$ is fixed on the boundary,
the interaction can only involve $X$; conformal invariance then
dictates
\bbb
  \delta\SS_{\rm bdy} &=& \beta\oint \cos(X)\qquad,\qquad (X~{\rm Euclidean})
\label{Xoptach}\\
  \delta\SS_{\rm bdy} &=& \cases{\beta\oint \cosh(X)& \cr
		\beta\oint\sinh(X) & } \ ,\qquad (X~{\rm Lorentzian})
\nonumber
\eee
This interaction is the boundary, open string analogue of the 
closed string tachyon background $V(X)$ in \pref{sigmod}; 
it describes an open string `tachyonic mode' of the D-brane,
since the interaction grows exponentially in Lorentz signature
spacetime.%
\footnote{The mass shell condition for boundary
interactions describing background fields on the D-brane
can be computed along the lines of
\pref{Tope}-\pref{onshell}, except that the OPE
of the stress tensor with a boundary operator should be
performed using the appropriate Dirichlet or Neumann propagator, 
$X(z)X(w)\sim -\frac{\alp}2(\log|z-w|^2\mp\log|z-\bar w|^2)$.}

The open string tachyon \pref{Xoptach} describes the decay
of an unstable D-particle located in the strong coupling
region $\phi\to\infty$.
The tachyon condensate in Lorentz signature
looks promising to be the description 
of an eigenvalue in the matrix model, whose
classical motion is
\bbb
  \lam(x) &=& \lam_0\,\cosh(x)\quad,\qquad E=-\coeff12\lam_0^2<0
\nonumber\\
  \lam(x) &=& \lam_0\,\sinh(x)\quad,\qquad E=+\coeff12\lam_0^2>0
\label{classtraj}
\eee
depending on whether the eigenvalue passes over, or is reflected by,
the harmonic barrier.  Similarly, the Euclidean trajectory 
$\lam(x) = \lam_0\cos(x)$
is oscillatory, appropriate to the computation of
the WKB tunnelling of eigenvalues under the barrier.

How do we see that this is so?   In
\cite{Douglas:2003up} 
(building on earlier work \cite{McGreevy:2003kb,Klebanov:2003km})
this result was demonstrated by computing the ground ring
charges $\OO_{12}$ and $\OO_{21}$ on the disk,
and showing that they give the classical motions above.
Here we will employ a complementary method: We will probe
the D-brane motion with the macroscopic loop.
This will exhibit the classical motion quite nicely.


\subsection{\label{lasso}Lassoing the D-particle}

The matrix model calculation of a macroscopic loop
probing a matrix eigenvalue is trivial.
Recall that the macroscopic loop is 
\be
  W(\lz,x_0) = -\frac1N\,\tr\log(\lz-M(x_0))
	= -\int d\lam\,\hat \rho(\lam,x_0)\,\log(\lz-\lam)\ .
\label{loopeval}
\ee
An individual eigenvalue undergoing classical motion
along the trajectory $\lam(x_0)$ gives a 
delta-function contribution to the eigenvalue density
\be
  \delta\!\rho(\lam,x_0) = \delta(\lam-\lam(x_0))
\label{deltarho}
\ee
where $\lam(x_0)=-\lam_0\cos(x_0)$ for Euclidean signature,
and $\lam(x_0)=-\lam_0\cosh(x_0)$ for Lorentzian signature.
Plugging into \pref{loopeval}, we find 
\be
  W_{\rm eval}(\lz,x_0) = -\log[\lz-\lam(x_0)]\ .
\label{Weval}
\ee

In the worldsheet formalism,
the presence of a macroscopic loop introduces a second boundary,
besides the one describing the D-particle.  The leading order
connected correlator of the loop and the D-particle is
thus an annulus amplitude; the worldsheet and
boundary conditions are depicted in 
figure \annulus, while the spacetime interpretation is
shown in figure \loopamps b.  The parameter $\tau$ is an example
of a modulus of the surface, the Schwinger parameter
for the propagation of a closed string, which cannot be gauged
away by either reparametrizations or local scale transformations;
in the end, we will have to integrate over it.

\begin{figure}[ht]
\begin{center}
\[
\mbox{\begin{picture}(240,150)(0,0)
\includegraphics[scale=.6]{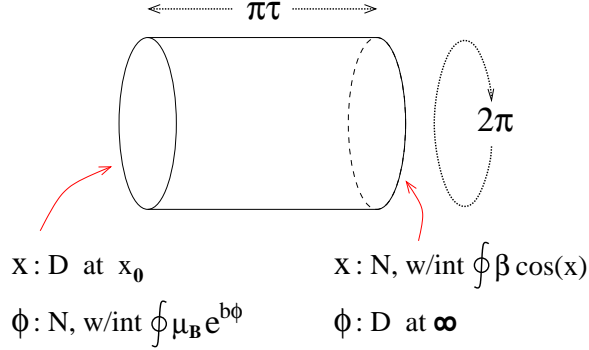}
\end{picture}}
\]
\caption{\it 
The annulus worldsheet describing a macroscopic loop probing
a D-particle.
}
\end{center}
\end{figure}

There are two ways to think about this worldsheet
as the propagation of a string.  If we view worldsheet
time as running around the circumference of the annulus,
we think of the diagram as the one-loop vacuum amplitude
of an open string, a string having endpoints.
At one endpoint of the string, we classically have
the boundary condition 
\be
  \partial_n\phi = 2\pi\mub\,e^{b\phi}\quad,\qquad X=x_0
\label{NphiDx}
\ee
describing the macroscopic loop;
at the other end, we have
\be
  \phi = \infty\quad,\qquad
  \partial_nX = 2\pi\beta\,\sin(X)
\label{DphiNx}
\ee
describing the moving D-particle.
On the other hand, we can think of the diagram as the
propagation of a closed string for a worldsheet time $\pi\tau$,
folded into ``boundary states''  $\ket{B}$ which implement
the boundary conditions on the fields.  These boundary
states are completely determined by these conditions, \eg\
\bbb
  & &(\partial_n\phi - 2\pi\mub\,e^{b\phi})\ket{B_N(\mub)}_\phi^{~} = 0
\nonumber\\
  & & (X-x_0)\ket{B_D(x_0)}_X^{~}=0
\label{Bstconds}
\eee
and so on.  Because the Liouville and matter fields do not interact,
the boundary state factorizes into the tensor product of
the boundary states for $X$ and for $\phi$.
The Liouville partition function can then be written
\bbb
  \ZZ_L(q) &=& \bra{B_N^{~},\mub} e^{-\pi\tau H} \ket{B_D^{~}}
\nonumber\\
	&=& \int d\nu\,\Psi^*_\fzzt(\nu,\mub)\Psi_\zz(\nu)
		\,\frac{q^{\nu^2}}{\eta(q)}
\label{liouann}
\eee
where $q=\exp[-2\pi\tau]$, and the Dedekind eta function
$\eta(q)=q^{1/24}\prod_{n=1}^\infty(1-q^n)$ represents
the contribution to the partition function of all the
Liouville oscillator modes.  The quantities
$\Psi_{\fzzt}$ and $\Psi_\zz$ are the zero mode parts
of the Neumann and Dirichlet boundary state wavefunctions,
respectively; see
\cite{Fateev:2000ik,Teschner:2000md}
and
\cite{Zamolodchikov:2001ah}, 
respectively.  Explicitly,
\bbb
  \Psi_\fzzt(\nu,\mub) &=& \cos(2\pi\nu s)\,\Bigl[
	\frac{\Gamma(1+2i\nu b)\Gamma(1+2i\nu/b)}%
		{2^{1/4}\,(-2\pi i\nu)}\; \muhat^{-i\nu/b}\Bigr]
\label{Lpsidef}\\
  \Psi_\zz(\nu) &=&
	2\,\sinh(2\pi\nu/b)\sinh(2\pi\nu b)\Bigl[
		\frac{\Gamma(1+2i\nu b)\Gamma(1+2i\nu/b)}%
		{2^{1/4}\,(-2\pi i\nu)}\; \muhat^{-i\nu/b}\Bigr]
\nonumber
\eee
Here, $s$ parametrizes $\mub$ as in equation \pref{mubparam}.
The ``Neumann'' wavefunction $\Psi_\fzzt$ is the one
obtained before, from the macroscopic loop calculation;
$\nu$ is the Liouville zero-mode momentum $\alpha=\coeff12 Q+i\nu$
in the ``closed string channel''.
This is not surprising; before we used the macroscopic loop
to probe the wavefunction of a scattering state,
now we are using it to probe a D-brane state to see
if it has the properties of a matrix eigenvalue.

The authors of
\cite{Zamolodchikov:2001ah}
showed that $\Psi_\zz(\nu)$ has the property that
all operators behave like the identity operator
as they approach the corresponding boundary
(so that one approaches the constant negative curvature
``vacuum'' near the boundary of the Poincar\'e disk).
Ordinarily in Liouville theory, 
when an operator such as $\V_\alpha$ approaches
the boundary $z=\bar z$ (\eg\ with boundary condition \pref{NphiDx}),
it expands as a sum of boundary operators $B_\beta$.  
For the boundary state with wavefunction $\Psi_\zz(\nu)$,
only the identity boundary operator 
$B_0=\One$ appears in the limit $z\to\bar z$.

Now for the matter partition function.  The annulus partition
function with the requisite boundary conditions
was computed in
\cite{Callan:1994ub}
for Euclidean $X$, with the result
\be
  \ZZ_X = \frac{1}{\sqrt2 \,\eta(q)}\sum_{n=-\infty}^{\infty}
	q^{n^2/4}\,\cos[n\pi(\coeff12+\gamma)]
	\quad,\qquad 
	\sin(\pi\gamma)\equiv\cos(x_0)\sin(\pi\beta)\ .
\label{Xann}
\ee
Again the Dedekind eta function represents the contribution
of the $X$ oscillator modes, and the sum results from
the zero modes.
The Faddeev-Popov ghost partition function is
\be
  \ZZ_{\rm gh} = \eta^2(q)\ ,
\label{Zgh}
\ee
cancelling the oscillator $\eta$ functions of $\phi$ and $X$.
This is related to the fact that there are no transverse directions
in which the string can oscillate -- at generic momenta
just the tachyon, with only center of mass motion 
of the string, is physical.

Combining all the contributions, we have
\bbb
  \ZZ &=& \int_0^\infty d\tau\,\ZZ_L\cdot\ZZ_X\cdot\ZZ_{\rm gh}
\nonumber\\
	&=& \int_0^\infty d\tau\int d\nu \cos(2\pi\nu s)
	\sum_{n=-\infty}^\infty \cos[n\pi(\coeff12+\gamma)]
		\, q^{\nu^2+n^2/4}\ .
\label{Zcombined}
\eee
Doing the $\tau$ integral, and the $\nu$ integral by residues,%
\footnote{The $n=0$ term is divergent and needs to be regularized.
Our choice is to replace the double pole at $\nu=0$ in this term by
$\frac1{\nu^2+\epsilon^2}$, and then subtract the pole term in 
$\epsilon$ after the $\nu$ integral is evaluated by residues.}
one finds
\be
  \ZZ = 2 \sum_{n=1}^\infty \frac 1n\,\exp[-n\pi s]
	\cos[n\pi(\coeff12+\gamma)]\ .
\label{Znext}
\ee
The sum is readily performed, and after a little algebra, one obtains
\be
  \ZZ = -\log[2(\cosh(\pi s) + \sin(\pi\gamma))]\quad .
\label{Zfinal}
\ee
Define now
\be
  \lam(x_0) = -\sqrt{2\mu}\,\sin(\pi\gamma)
	= -\sqrt{2\mu}\,\sin(\pi\beta)\,\cos(x_0)
	\equiv -\lam_0\cos(x_0)
\label{clastrj}
\ee
and recall that $\mub=\sqrt{2\mu}\,\ch(\pi s)=\lz$; then we have
\be
  \ZZ = -\log[\lz-\lam(x_0)]+\coeff12\log(\mu/2)\ .
\label{tadaa}
\ee
The additive constant is ambiguous, and depends on how we regularize
the divergent term in \pref{Zcombined}; nevertheless,
it is independent of the boundary data for $X$ and $\phi$,
and so does not affect the measurement of the D-particle motion.%
\footnote{The extra term
$(\frac1\epsilon-\frac12\log(\mub))$ can be thought of as the
regularized volume of $\phi$-space \cite{Kutasov:2004fg}.} 
Dropping this last term, we finally reproduce
the result (so easily found) for the probe
of eigenvalue motion in the matrix model, equation \pref{Weval}!

Note that, even though the Dirichlet boundary condition 
on $\phi$ is in the strong coupling region $\phi\to\infty$,
the wavefunctions are such that we obtain sensible
results for the amplitude.
Note also that the boundary interaction for the D-particle
depends only on $X$, and thus the D-particle naively
is not moving in $\phi$.  This is a cautionary tale,
whose moral is to compute physical observables!
Nevertheless, when the D-particle reaches the
asymptotic region of weak coupling, it should be moving in both
$\phi$ and $\lambda$.  Somehow the field space coordinate of
the open string tachyon and the $\phi$ coordinate of spacetime
become related in the course of the tachyon's condensation,
and it remains to be understood how this occurs.

The consideration of multiple D-particles elicits the matrix
nature of their open string dynamics.  Now we must add to
the description of the boundary state a finite dimensional
({\it Chan-Paton}) Hilbert space $\HH_{\cp}$ describing 
which D-particle a given worldsheet boundary is attached to.
Open string operators act as operators on this 
finite-dimensional Hilbert space, \ie\ they are matrix-valued
(equivalently, an open string is an element of 
$\HH_\cp\otimes\HH_\cp^*$ specifying the Chan-Paton 
boundary conditions at each end).

The open string tachyon is now a matrix field;
the parameter $\beta$ in equation \pref{Xoptach}
is a matrix of couplings $\beta_{ij}$, $i,j=1,...,n$
for $n$ D-particles.  There is an additional possible
boundary interaction
\be
  \delta\SS_{\rm bdy} = \oint A_{ij}\,\partial_t X
\label{opgauge}
\ee
which is a matrix gauge field on the collection of D-particles.  
We ignored it in our previous discussion because its role
is to implement Gauss' law on the collection of D-particles,
which is trivial in the case of a single D-particle.
When several D-particles are present, however, this Gauss
law amounts to a projection onto $U(n)$ singlet states.
Thus the continuum description suggests that the $U(N)$
symmetry of the matrix mechanics is gauged, which
as mentioned in section \ref{nonsinglet}
projects the theory onto $U(N)$ singlet wavefunctions.
Singlet sector matrix mechanics looks very much like
the quantum mechanics of $N$ D-particles in 2D string theory.

Several ingredients of the relation between the continuum
and matrix formulations remain to be understood.
The probe calculation tells us that the open string
tachyon condensate on the D-particle
describes its leading order, classical trajectory.
One should understand how higher order corrections
lead to the quantum corrections for the wavefunction of a quantum
D-particle, and show that this series matches the 
WKB series for the wavefunctions of the eigenvalue fermions
of the matrix model.  Also, the description in the continuum
formulation of an eigenvalue as a D-particle is quite different
from the ensemble of eigenvalues in the Fermi sea, whose
collective dynamics is expressed via continuum worldsheets.
Under what circumstances is eigenvalue dynamics that of D-particles,
as opposed to that of closed string worldsheets?
For instance, the $U(n)$ symmetry of a collection of $n$ D-particles
should extend to the full $U(N)$ symmetry of the whole matrix.
How do we see this larger symmetry in the continuum formulation?
We cannot simply turn all the fermions into D-particles;
there would then be nothing left to make the continuum worldsheets 
that attach to these branes.  The continuum formalism is really
adapted to describing a small number of matrix eigenvalues
that have been separated from rest of the ensemble,
thus leading to distinct treatment of the few separated ones
as D-particles, and the vast ensemble of remaining ones
as the threads from which continuum worldsheets are woven.

\subsection{Summary}

To summarize, we have an expanding translation table between
matrix and continuum formalisms:

\vskip .2cm
\begin{center}
\begin{tabular}{||c|c||}
\hline
{\bf Continuum} & {\bf Matrix} \\ \hline\hline
Closed string vacuum & 
	Fermi sea of matrix eigenvalues \\ \hline
Liouville potential $\mu\, e^{2b\phi}$ & 
	Inverted oscillator barrier \\ \hline
Worldsheet cosmological constant $\mu$ &
	Fermi energy $-2\mu$ \\ \hline
Strings &
	Fermi surface density wave quanta \\ \hline
String S-matrix &
	Density wave S-matrix \\ \hline  
D-branes w/$X$: D, $\phi$: N, $\SS_{\rm bdy}\!=\!\mub\oint\! e^{b\phi}$ &
	Macroscopic loops $\tr[\log(\lz-M(x))]$ \\ \hline
Boundary cosm. const. $\mub$ &
	Loop eigenvalue parameter $2\lz$ \\ \hline
D-branes w/$\phi$: D, $X$: N, $\SS_{\rm bdy}\!=\!\beta\oint\!\cos X$ &
	Eigenvalues outside the Fermi sea \\ \hline
Open string tachyon coupling $\beta$ &
	Eigenvalue energy $E=-\mu\,\sin^2\pi\beta$ \\ \hline
Open string tachyon on $n$ D-particles &
	A block of the matrix $M$ \\ \hline
\end{tabular}
\end{center}
\vskip .2cm

A similar dictionary is known for the fermionic string,
which will be described briefly in the next section.
Here one has the added advantage that the model
is nonperturbatively well-defined.
In these models both
sides of the duality are again calculable.  
One may hope that open/closed string duality can be worked out in
complete detail in this example, and that it will lead
to valuable insights into the general class of open/closed
string dualities to which it belongs.


\section{\label{fermstr}Further results}

\subsection{Fermionic strings}

The remarkable agreement between the continuum and matrix formulations
of 2D string theory leads us to believe that they are equivalent.
However, in the case of the bosonic string, both are asymptotic
expansions.  Worldsheet perturbation theory is an asymptotic expansion,
and it was our hope that, as in higher dimensional gauge/gravity
correspondences, the matrix (gauge) theory formulation would 
provide a nonperturbative definition of the theory.  But the
nonperturbative instability of the vacuum to eigenvalue tunnelling
across to the right-hand side of the oscillator barrier means that
the theory does not really exist after all.

An obvious fix for this difficulty would be to fill up
the other side of the barrier with fermions as well
(see \eg\ \cite{Dhar:1995gw} for an example of this proposal).
But this leads to an equally obvious question: We found
agreement with continuum bosonic strings using just the
fluctuations on one side of the oscillator barrier.
What do the fluctuations on the other side describe?
Perturbatively, they are a second, decoupled copy of the same dynamics.
Nonperturbatively, the two sides of the barrier communicate,
by tunnelling and by high energy processes that pass over
the barrier.  It is now understood 
\cite{Takayanagi:2003sm,Douglas:2003up}
that this stable version of the matrix model describes
2D fermionic string theory (the type 0B string, in the arcane
terminology of the subject).

The fermionic string extends the construction of section \ref{strth}
by supersymmetrizing the worldsheet theory: The spacetime coordinates
$X^\mu(\sigma)$ of the worldsheet path integral gain superpartners
$\psi^\mu(\sigma)$ which transform worldsheet spinors 
(and spacetime vectors).  The local reparametrization and
scale invariance condition generalizes to local supersymmetry
and super-scale invariance; in other words, the stress tensor
$T$ has a superpartner $G$, and the dynamical condition is
that both must vanish in correlation functions.

If we perform the same exercise in the path integral formulation
\pref{ptclprop} of the particle propagator
in flat spacetime, the quantization
of the superpartner $\psi$ leads to equal time anticommutation relations
\be
  \{\psi^\mu,\psi^\nu\} = \delta^{\mu\nu}\ .
\label{psiccr}
\ee
One realization of these anticommutation relations
is to represent the $\psi$'s as Dirac matrices.
The quantum mechanical Hilbert space contains not
only the position wavefunction, but also a finite
dimensional spin space in which the $\psi^\mu$ act --
the particle being propagated is a spinor.  
Thus worldsheet supersymmetry is a way to introduce spacetime fermions
into a worldline or worldsheet formalism.%
\footnote{Note that the wave operator is
the Hamiltonian for the quantum mechanics,
\eg\ $H=-p^2+m^2$ in flat spacetime; the supersymmetry algebra
$Q^2=H$ points to the fact that the supercharge 
$Q=p\cdot\psi$ is the Dirac operator in this context.}
A second realization of the anticommutation relations
\pref{psiccr}, using complex fermions,
treats $\psi^*_\mu$ as a creation operator,
and its conjugate $\psi^\mu$ as an annihilation operator.
Starting from the fermion `vacuum' $\ket{0}$,
$\psi^\mu\ket{0}=0$, the set of polarization states
propagated along the particle worldline,
$\{\psi^*_{\mu_1}\cdots\psi^*_{\mu_r}\ket{0}\}$,
transform as a collection of antisymmetric tensors 
$C_{\mu_1...\mu_r}$ in spacetime.

The same story arises in the string generalization;
the worldsheet fermions $\psi^\mu$ can realize a collection of 
antisymmetric tensors in spacetime, or under suitable conditions
the propagating strings are spinors in spacetime.
The so-called type 0 fermionic strings do not 
realize spacetime fermions, but do contain the
antisymmetric tensor fields.  We can divide the
set of antisymmetric tensor fields into those with
even rank and those with odd rank.  The type 0A theory
involves a projection onto odd rank tensors
(with even rank field strength), while the type 0B
theory contains even rank tensors (with odd rank
field strength).  In particular, the type 0B theory
contains a 0-form or scalar potential $C$ in addition
to the tachyon $V$.  This scalar provides the
needed extra degrees of freedom to represent, 
in the worldsheet formalism, the density oscillations
on either side of the harmonic barrier in the 
matrix model with both sides filled.

To describe the vertex operator for this scalar
requires a bit of technology \cite{Friedan:1985ge}.  
One can think of the left- and right-moving worldsheet fermions 
$\psi(z)$, $\bar\psi(\bar z)$ in terms of 
the 2d Ising model.  In addition to the fermion
operators, the Ising model has order and disorder operators 
$\sigma(z,\bar z)$ and $\mu(z,\bar z)$,
often called {\it spin fields}.
In 2D string theory, we thus have the spacetime coordinate fields
$X$, $\phi$ and their superpartners $\psi_{\sx}$, $\psi_{\sphi}$,
as well as the spin fields $\sigma_{\sx}$, $\sigma_{\sphi}$,
$\mu_\sx$, $\mu_\sphi$, as the ingredients out of which
we can build vertex operators (there may also be
contributions from Faddeev-Popov ghosts if this is required
by gauge invariance).  The on-shell tachyon vertex is now
\be
  V_{i\omega} = (\psi_\sx\pm\psi_\sphi)(\bar\psi_\sx\pm\bar\psi_\sphi)
		\,e^{i\omega(X\pm \phi)}\,e^{Q\phi}\ ,
\label{tztach}
\ee
and there is also the second (so-called {\it RR}) scalar,
with vertex operator
\be
  C_{i\omega} = \Sigma_{\rm gh}(\sigma_\sx\sigma_\sphi\pm\mu_\sx\mu_\sphi)
		\,e^{i\omega(X\pm \phi)}\,e^{Q\phi}\ .
\label{pierre}
\ee
Here, $\Sigma_{\rm gh}$ is a spin field for the Faddeev-Popov
ghosts arising from fixing local supersymmetry \cite{Friedan:1985ge}.
It was shown in \cite{DiFrancesco:1991ud} 
that the tree-level S-matrix amplitudes
for the linear combinations
\be
  T_{\sst L,R}(\omega) = 
	\coeff{\Gamma(-i\omega\sqrt{\alp/2})}{\Gamma(i\omega\sqrt{\alp/2})}
	\, V_{i\omega}\,
	\pm\,\coeff{\Gamma(\frac12-i\omega\sqrt{\alp/2})}%
		{\Gamma(\frac12+i\omega\sqrt{\alp/2})}
	\, C_{i\omega}
\label{LRmodes}
\ee 
decouple from one another, \ie\ the connected amplitudes involving
both sets of operators $T_{\sst L,R}$ vanish; and the amplitudes 
involving just one set are the same as for the bosonic string,
up to a rescaling $\alp\to 2\alp$.  This strongly suggests
we identify $T_{\sst L,R}$ as the asymptotic modes of
density fluctuations on the left and right sides of the
harmonic barrier in the symmetrically filled matrix model.
Note that there are again energy-dependent phases
involved in the relation between matrix model asymptotic states
and continuum asymptotic states.  One should think of
the fields $V$ and $C$ as corresponding to 
the symmetric and antisymmetric perturbations of the 
Fermi sea of the matrix model, after these phases
are stripped off.  This identification is consistent with
the fact that S-matrix amplitudes vanish for an odd
number of parity-odd density perturbations; the $Z_2$ Ising
symmetry causes the correlator of an odd number of spin fields 
to vanish as well.
The two-to-one map of $\lam$-space to $\phi$-space in
the type 0B model highlights their nonlocal relation,
a feature we have already seen several times.

This proposal passes checks analogous to the bosonic string --
the tree level S-matrix, the torus partition function,
and expectations of the ground ring operators on the sphere
and on the disk, all agree between matrix and continuum approaches
\cite{Douglas:2003up}.

The $\IZ_2$ symmetry that changes the sign
of spin operators like $C_{i\omega}$,
called {\it NSR parity}, also characterizes the boundary
states, splitting them into $\IZ_2$ even ({\it NS}) 
and odd ({\it R}) components.  For instance, there are 
separate {\it NS} and {\it R} macroscopic loops.
We may determine their functional form 
in the matrix model by repeating
the calculation of section \ref{lasso}.
The main differences will be that the calculation splits
into these two boundary state sectors.
The boundary state wavefunctions $\Psi_{\!NS}$ and $\Psi_{\!R}$
for both Dirichlet (ZZ) and Neumann (FZZT) branes 
appearing in \pref{liouann} are given in 
\cite{Fukuda:2002bv,Ahn:2002ev}
(see also \cite{Douglas:2003up}, sections 6 and 7).  
The matter partition function on the annulus
\cite{Gaberdiel:2001zq}
is essentially the same as equation \pref{Xann},
with the sum over $n$ restricted to even integers
in the NS sector and odd integers in the R sector.
The analysis then proceeds along the lines of section \ref{lasso};
one finds
\bbb
  \ZZ_{\rm\sst NS} &=& -\hf\log[\mu_{\!B}^2-\lam^2(x_0)]+\hf\log(\mu/2)
\nonumber\\
  \ZZ_{\rm\sst R} &=& 
	\hf\log\Bigl[\frac{\mub-\lam(x_0)}{\mub+\lam(x_0)}\Bigr]\ .
\label{fermann}
\eee
These results prove a conjecture 
\cite{Takayanagi:2003sm,Takayanagi:2004jz}
for the form of the macroscopic loop operators in the matrix model
for the type 0B fermionic string.

The super-Liouville boundary state wavefunctions 
\cite{Fukuda:2002bv,Ahn:2002ev}
are also the major ingredients of the disk one-point functions
that yield the wavefunctions corresponding to the operators
$V_{i\omega}$ and $C_{i\omega}$.  For the tachyon, 
one finds essentially the same result \pref{contonept},
while for the RR scalar $C$, one finds 
\be
  \vev{C_{i\omega}(z,\bar z)}_{\rm disk} = \frac{1}{2\pi}
	\, \muhat^{-i\omega/2}
	(\Gamma(\coeff12+i\omega))^2\,\cos(\pi s\omega)\ .
\label{RRonept}
\ee
The corresponding integral transforms to loop length wavefunctions
again yield Bessel functions
\cite{Takayanagi:2003sm,Douglas:2003up,Takayanagi:2004jz}.

There are also a few discrete symmetries that match
on both sides of the correspondence.  
One example is the $\lam\to-\lam$ parity 
symmetry of the matrix model, which appears as the 
$\IZ_2$ NSR parity symmetry which sends $C\to-C$ in the continuum theory.
The continuum theory also has a symmetry under
$\mu\to-\mu$;%
\footnote{Combined with $\psi_{X,\phi}^{~}\to -\psi_{X,\phi}^{~}$
(but keeping $\bar\psi_{X,\phi}^{~}$ unchanged).
This symmetry is the discrete $\IZ_2$ {\it R-parity} symmetry
of the worldsheet supersymmetry of the model.}
in the matrix model, this is the symmetry
of the Hamiltonian $H=\half(p^2-\lam^2)$ under $p\leftrightarrow\lam$,
combined with an interchange of particles and holes.
A few other checks, as well as a second
2D fermionic string model -- the type 0A string, whose matrix model
formulation involves the dynamics of open strings in
a system of D-particles and their antiparticles -- can be found in 
\cite{Douglas:2003up}.

\subsection{Remarks on tachyon condensation}

The structure of the bosonic and fermionic matrix models of
2D string theory is a remarkable illustration 
of the effective picture of tachyon condensation 
on systems of unstable D-branes
\cite{Sen:1999mg}.
In perturbative string theory, a D-brane is a heavy,
semiclassical object much like a soliton.
The analogy to solitons is in fact quite precise
\cite{Witten:1998cd,Sen:1999mh,Harvey:2000tv,Harvey:2000jt}.
Unstable D-branes are like solitons
that do not carry a topological charge, and thus
can decay to the vacuum (plus radiation).  But being heavy,
the initial stages of the decay are a collective process
of instability of the `soliton' field configuration.  
The quanta of this unstable mode are open string
tachyons, and the initial stages of the decay
are best described as the condensation of this
tachyonic mode.  An effective potential picture
of this process is shown in figure \senpotl a
for an unstable brane in the bosonic string.

\begin{figure}[ht]
\begin{center}
\[
\mbox{\begin{picture}(348,115)(0,0)
\includegraphics[scale=.5]{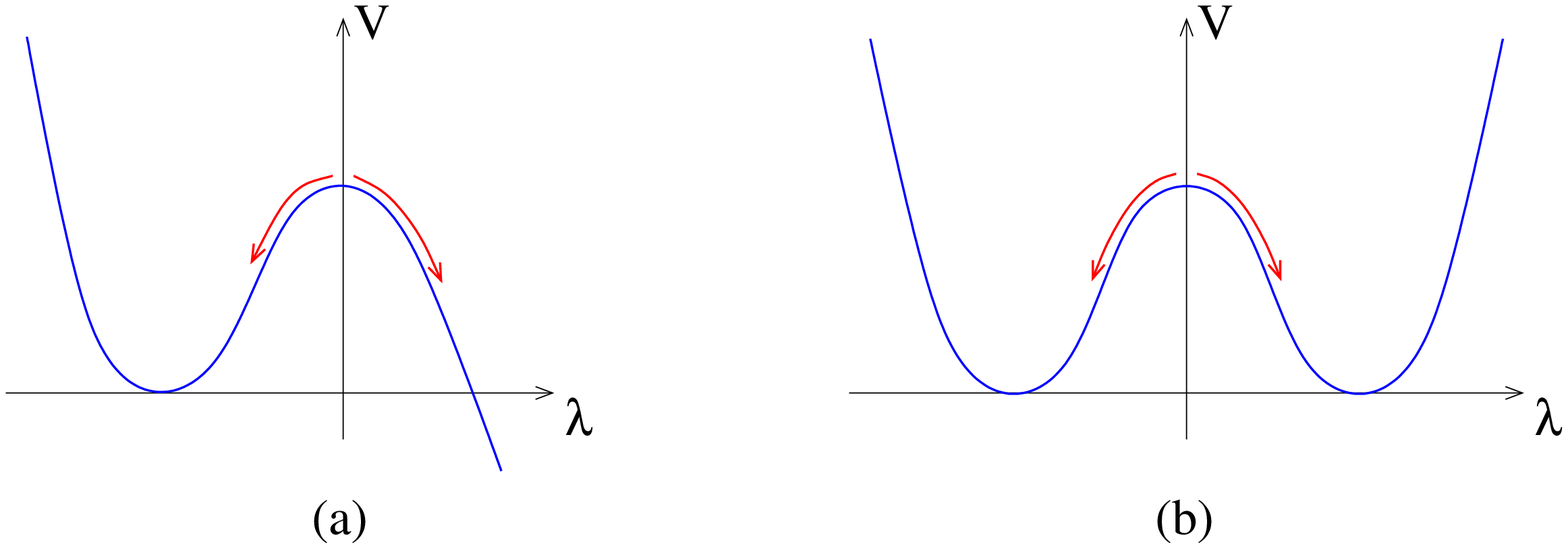}
\end{picture}}
\]
\caption{\it 
The effective potential for the open string tachyon
on an unstable D-brane in the (a) bosonic, and
(b) fermionic strings.
}
\end{center}
\end{figure}

The heuristic picture of the effective potential identifies
the local minimum to the left of the unstable point
with the ``closed string vacuum'', and the difference
in energy between local maximum and local minimum
is the energy of the initial unstable brane.
An initial state of the tachyon field $T$
localized at the unstable maximum of the effective
potential is meant to describe the presence of
the unstable brane, and condensation of $T$
describes its decay.  Condensation to $\vev{T}<0$
represents decay toward the closed string vacuum.
The abyss to the right of the local maximum 
is meant to represent the fact that condensing the
open string tachyon to $\vev{T}>0$
leads to singularities at finite time 
in perturbative calculations 
\cite{Sen:2002nu} with no known
string interpretation; it is not understood whether there
is any stable, nonsingular state to which the system evolves
when the open string tachyon condenses in that direction.

Qualitatively, this picture is identical to that of 
the matrix potential of figure \cubicpotl a.
The only difference is that the closed string vacuum
is itself described via the open-closed string equivalence
as a degenerate gas of D-particles -- in a sense a collection
of unstable D-branes that have ``already decayed''.
The absence of eigenvalues to the right of the barrier
means that there is no worldsheet interpretation
for eigenvalues in this region, just like the region
to the right of \senpotl a.

A similar story applies to the fermionic string.
The open string tachyon
effective potential has two symmetric wells,
as in figure \senpotl b.
On one hand, condensation in either direction
of the open string tachyon
on an unstable D-brane leads to its decay to the
closed string vacuum; on the other hand, the matrix model
for the type 0B string has just such a potential,
with both wells filled by eigenvalue fermions,
and a string interpretation of the physics on
either side.  The analogy also holds for the matrix
model equivalent of the type 0A string.

\section{Open problems}

What remains to be understood?  
In this concluding section, let us list a few 
unresolved issues and directions for future work.


\subsection{The open-closed string duality map}

While there is a qualitative
map \pref{Wrep} between $\lam$-space and $\phi$-space
at the level of zero modes, a precise map between
the matrix model and the full Liouville field theory
remains to be worked out.  This would require 
a complete translation between quantities in the matrix
model and the Liouville (plus free scalar) field theory.
One indication of a missing ingredient is that the
asymptotic states of the matrix model have the 
leg poles of the continuum formalism stripped off,
see equation \pref{paramrels}.  The poles incorporate
the effects of discrete physical states in the continuum
formulation \cite{DiFrancesco:1991ud}.  While, as argued above,
only center-of-mass string motion is physical at generic, continuous
momenta, there is an additional discrete spectrum of
physical states at special momenta
\cite{Lian:1991ty,Bouwknegt:1991yg,Witten:1992yj,Witten:1991zd};
a simple example is the zero-momentum
graviton vertex operator $V_{\rm\sst grav}=\partial X\bar\partial X$,
which is manifestly physical since it is the action density for $X$.
The continuum formalism knows how to incorporate gravitational effects,
while these are currently put into the matrix model by hand;
the matrix prescription for the S-matrix is to compute the LSZ-reduced
density wave scattering amplitudes, and then multiply the result
by a leg-pole factor $\frac{\Gamma(i\omega)}{\Gamma(-i\omega)}$
for each asymptotic state. 
It is this leg-pole factor which is responsible for perturbative
gravitational effects \cite{Natsuume:1994sp,DiFrancesco:1991ss}.


\subsection{Gravitational effects}

Perhaps a part of the explanation for this absence of
gravitational and other discrete state effects
in the matrix model is that, since the
linear dilaton lifts the string ``tachyon'' mode to zero mass,
it also raises the graviton to positive mass; its effects
are subleading to the tachyon, and might be masked by or effectively
absorbed into tachyon dynamics
\cite{Banks:1991sg,Tseytlin:1991bu}.

Initially there was hope that the matrix model
would teach us about nonperturbative gravity,
and in particular lead to a solvable model
of black hole dynamics. 
A second background solution to the string equations
of motion \pref{dynprinc}-\pref{betafns} appears to be a black hole
\cite{Witten:1991yr,Mandal:1991tz}
\bbb
  ds^2 &=& d\phi^2 \pm \tanh^2(\coeff{1}{\sqrt{2(k-2)}}\,\phi)\,dx^2
\nonumber\\
  \Phi &=& \Phi_0 + \log[\cosh(\coeff{1}{\sqrt{2(k-2)}}\,\phi)]\ ,
\label{twodbh}
\eee
depending on whether we are interested
in Euclidean or Lorentzian signature.
The metric is written in Schwarzschild-like coordinates,
where the horizon at $\phi=0$ is infinitely redshifted
relative to the asymptotic region $|\phi|\to\infty$.
This sigma model \pref{sigmod} on this background 
describes an exact conformal field theory,
the $SL(2,\IR)/U(1)$ gauged WZW model
(the signature is determined by the conjugacy class
of the $U(1)\subset SL(2,\IR)$ being gauged). 
The level $k$ of the $SL(2,\IR)$ current algebra
symmetry of the WZW model
is $k=9/4$ for the bosonic string, and $k=5/2$ for 
the fermionic string, in order that the slope $Q=\sqrt{\frac{2}{k-2}}$
of the asymptotically linear dilaton in \pref{twodbh}
have the right value for 2D string theory.
Note that the radius of curvature of the geometry is of order $1/\sqrt k$
in the vicinity of the horizon $\phi\sim 0$; therefore
it is important to have an exact conformal field theory,
since the corrections to the leading
order equations of motion \pref{betafns} are significant.
Note also that the leading asymptotic perturbation 
$e^{2Q\phi}\partial X\bar\partial X$
of the metric away from flat spacetime,
is the reflected version (the other on-shell value
of Liouville momentum) of the special physical
graviton operator $\partial X\bar\partial X$ discussed above.
Thus the background can be thought of as the nonlinear
completion of this linearized deformation.
A shift in $\phi$ makes $e^{-2\Phi_0}$ the coupling in front
of the asymptotic graviton in \pref{twodbh};
as in higher dimensions, the coefficient
of the leading asymptotic deformation
of the metric away from flat spacetime
is the mass of the black hole
\cite{Gibbons:1992rh,Nappi:1992as},
$\mu^{~}_{\rm bh}= e^{-2\Phi_0}$.

A great deal is known about this CFT.
There is a conformal bootstrap, analogous to
that of Liouville theory 
\cite{Teschner:1997ft,Giveon:2001up}.
The analogue of the two degenerate operators
$V_{-b/2}$, $V_{-1/2b}$ of Liouville theory are the
degenerate operators $\Phi_{j}$
of $SL(2,\IR)$ current algebra, 
having spin $j=-\frac32$ and $j=-\frac{k}2$.

A rather remarkable conjecture \cite{FZZduality}
claims that the Euclidean $SL(2,\IR)/U(1)$ gauged WZW model
is equivalent as a quantum field theory to
another model, the so-called Sine-Liouville theory,
whose action is
\be
  \SS_{\rm SL} = \frac{1}{4\pi} \int \sqrt g\,
	\Bigl[g^{ab}\partial_a\phi\partial_b\phi
	+ g^{ab}\partial_a X\partial_b X
		+QR^{\sst (2)}\phi
	+\mu_{\rm sl}^{~}\cos R[X_l-X_r]\,e^{\frac1Q\phi}\ .
\label{SLact}
\ee
Here $Q^2=\frac{2}{k-2}$; $X$ is compactified on a circle of radius
$R=2/Q$; and $X_l-X_r$ is the axial component of $X$,
so that the potential in \pref{SLact}
acts as a generating function for vortices in the 
worldsheet partition function.
In \cite{Giveon:2001up}, this equivalence is argued
to hold at the level of the conformal bootstrap
for correlation functions.
The ``resonant amplitudes'',
which are those correlators dominated in the path integral
by the asymptotic region $\phi\to-\infty$, 
involve only the operators 
$\Phi_{-\frac r2-\frac s2(k-2)-1}$, $r,s=1,2,...$.
These correlators must in general be perturbatively dressed by 
{\it both} the asymptotic graviton
$\mu_{\rm bh}^{~}\,e^{2Q\phi}\partial X\bar\partial X$,  
which dresses $r$,
and by the Sine-Liouville interaction 
$\mu_{\rm sl}^{~}\,e^{\frac1Q\phi}\cos R(X_l-X_r)$,
which dresses $s$.  
Self-consistency requires the coefficients of these two interactions
to be related \cite{Giveon:2001up};
one finds \cite{Giveon:2001up}
\be
  \pi \mu^{~}_{\rm bh}\, \frac{\Gamma(-Q^2/2)}{\Gamma(1+Q^2/2)}
        = \Bigl(\pi\mu_{\rm sl}^{~} \,Q^2/2\Bigr)^{Q^2}\ .
\label{bhslrel}
\ee
Again, as in Liouville theory there is a sense in which 
both dressing operators are present in the theory.

There is again a kind of strong/weak coupling duality,
since the metric deformation is dominant at weak coupling 
($\phi\to -\infty$) for $Q\ll 1$, 
while the Sine-Liouville coupling is dominant
for $Q\gg 1$.  Since $Q=2$ for the 2D string, 
one has the sense that the Sine-Liouville description 
is somewhat more appropriate.
In higher dimensions, when the curvature of a black hole
reaches string scale, it undergoes a phase transition 
to a gas of strings \cite{Horowitz:1996nw}
(the transition point is known as the {\it correspondence point}).
The apparent dominance of the Sine-Liouville coupling
may be an indication that the ``black hole'' of
2D string theory is actually on this other side
of the correspondence point, where it is better thought
of as a gas or condensate of strings.

The equivalence with Sine-Liouville leads to a natural candidate 
\cite{Kazakov:2000pm}
for a matrix model equivalent to the Euclidean
``black hole'' -- simply turn off the Liouville potential
and turn on a condensate of vortices in the
compactified Euclidean theory, \cf\ section \ref{nonsinglet}.
The matrix description of the background thus has a closer affinity
to the tachyon condensate of \pref{SLact} 
than it has to the Euclidean black hole of \pref{twodbh}.%
\footnote{After T-duality ($X_r\to -X_r$) and Wick rotation 
to Lorentz signature, the background \pref{SLact}
describes a moving tachyon wall 
$V_{\rm\sst backgd} = \mu_{\rm sl}\, e^{\phi/Q}\cosh(RX)$.  
Such time-dependent tachyon condensates have been considered in
\cite{Alexandrov:2002fh,Karczmarek:2003pv}.}

Yet another reason to suspect the absence of objects
that could truly be characterized as black holes in 
2D string theory, is the absence of nonsinglet states
in the matrix model.  As mentioned in the introduction,
the appearance of black holes in the density of states
in higher dimensional versions of the gauge/gravity
equivalence is associated to a deconfinement transition.
The thermodynamics one is led to 
\cite{Gibbons:1992rh,Nappi:1992as} 
on the basis of the classical gravity solution \pref{twodbh}
yields a density of states $\rho=\exp[\sqrt{2}\,\pi E]$.
Such a density of states will not come from the 
quantum mechanics of the degenerate Fermi gas
of the singlet secctor of the matrix model,
but might concievably come from 
the liberation of nonsinglet degrees of freedom of the matrix.
However, this is absent from the matrix model -- the $U(N)$
degrees of freedom are gauged away.

Indeed, a calculation \cite{Martinec:2004qt}
of nonperturbative high energy scattering in the matrix model -- 
a process that in higher dimensions
would certainly lead to the formation of black holes as 
long-lived intermediate states -- reveals none of the features
that would be predicted on the basis of the appearance
of black holes being formed during the scattering process.

In short, low energy gravitational effects are put into
the matrix model by hand, via the leg-pole factors.
High energy gravitational effects such as black hole
formation seem to be absent altogether.
Does the matrix model incorporate any form of 2d gravity?
If so, how?  If not, why not?

\subsection{Short-distance physics}

Even though it would appear that black hole physics is absent
from the matrix model, intriguing remnants of Planck scale
(or more precisely, ultra-short distance) physics seem to be present.  
Namely, the spacing of eigenvalues in the matrix model 
is of order the D-particle Compton wavelength 
$L_{\rm c}\sim e^\Phi \lstr$.

The fact that loop length scales as $\ell\sim e^{b\phi}$,
together with the integral transform \pref{Wrep},
suggests that the eigenvalue coordinate scales as $\lam\sim -e^{-b\phi}$
(in the sense of KPZ scaling).  
From equations \pref{fermop} and \pref{parcylas}
one determines $\vev{\hat\rho}\sim |\lam|$
as $\lam\to-\infty$.  The eigenvalue spacing
is $\delta\lambda\sim 1/\vev{\hat\rho}$, and thus
$\delta\lam/\lam\sim \lam^{-2}$.  In terms of the Liouville
coordinate, this spacing is $\delta\phi\sim e^{2\phi}=e^{\Phi}$,
which is $L_{\rm c}$!  This result generalizes to 
the discrete series of $c<1$ conformal field theories
coupled to Liouville gravity, which are thought
of as string theory in $D<2$.  Here we have $b=\sqrt{q/p}$,
with $p,q\in\IZ$ and $q<p$.  The pair $(p,q)$ characterize
the matter conformal field theory, with 
$c_{\rm matter}=1-6\coeff{(p-q)^2}{pq}$.
These models have a realization as an integral over two random matrices 
\cite{Tada:1991pa,Daul:1993bg,Kostov:1991cg}
with the eigenvalue density scaling as $\rho(\lam)\sim \lam^{p/q}$.
Tracing through the KPZ scaling, one finds
$\delta\lam/\lam\sim \lam^{-(1+1/b^2)}$, 
and once again the eigenvalue spacing is $\delta\phi\sim e^{Q\phi}=e^\Phi$.
An appealing interpretation of this result is that 
spacetime has a graininess or discrete structure at the
short distance scale $L_{\rm c}$. 
It would be interesting to find some `experimental'
manifestation of this spacetime graininess.

\subsection{Open string tachyons}

In higher-dimensional spacetime, the canonical picture
of the decay of unstable D-branes has the initial stages
of the decay well-described by open string tachyon
condensation; at late times the brane has decayed,
open strings are absent, and the energy is carried off
by a pulse of closed string radiation.

The qualitative picture is rather different in 2D
string theory.  Here the branes don't really decay;
the open string tachyon merely describes their motion in spacetime,
and there is an {\it equivalence} between two 
characterizations of the dynamics in terms of open
or of closed strings.

The worldsheet formulation has elements 
of both open and closed string descriptions of D-brane decay.
Closed string worldsheets represent the collective dynamics
of the Fermi sea of ``decayed'' eigenvalues; eigenvalues extracted 
from the sea are represented as D-branes with explicit
open string degrees of freedom.  
Thus the continuum description is naively overcomplete.
For instance, one can compute the ``radiation'' of closed strings from
the ``decaying'' D-brane representing an eigenvalue rolling
off the potential barrier
\cite{McGreevy:2003kb,Klebanov:2003km}.
One finds a closed string state
\be
  \ket{\psi} \sim \exp\Bigl[i\!\int\!{d\omega}\,
	v_p\, \alpha_p^\dagger\Bigr]\ket{{\rm vac}}
\label{dzerodecay}
\ee
with the coefficient $v_p$ given to leading order
by the disk expectation value of the tachyon vertex
operator $V_{i\omega}$ with the boundary conditions
\pref{DphiNx}.  Roughly, the closed string tachyon
bosonizes the eigenvalue fermion.  

Of course, the eigenvalue doesn't decay, but stays in its
wavepacket as it propagates to infinity.
An eigenvalue fermion maintains its identity
as it rolls to infinity; we are not forced
to bosonize it.  There appears to be some redundancy
in the worldsheet description, unless different descriptions
are valid in non-overlapping regimes (as is the case
in other open/closed string equivalences); but then
it remains to be seen what effects force us to 
describe the dynamics as that of D-branes or that of closed strings,
and in what regimes those effects are important.
A possible clue is the form of the D-brane boundary 
state, which fixes the boundary at $\phi=\infty$ 
throughout the motion, and instead describes the dynamics
as occuring in the field space of the open string tachyon.
On the other hand, we know that $\lambda\to -\infty$
corresponds to $\phi\to -\infty$, and therefore at late times
an appropriate boundary state should have significant support
in this weak coupling region.  This suggests that the
perturbative boundary state description of the rolling eigenvalue
breaks down at finite time.%
\footnote{I thank Joanna Karczmarek for discussions of this issue.} 
It was pointed out in \cite{Okuda:2002yd} that the boundary
state represents a source for closed strings that
grows exponentially in time, so that one would expect the
perturbative formalism to break down at a time of order
$x\sim \log\mu$ (note that this is roughly the WKB time of flight
from the top of the potential to the edge of the Fermi sea).
Once again we run into the issues surrounding
the quantization of the D-brane motion mentioned
at the end of section \ref{lasso}.

A similar issue is the absence so far of a completely convincing
worldsheet description of holes in the Fermi sea of eigenvalues
(for a proposal based on analytic continuation
of the boundary states, see \cite{Douglas:2003up,Gaiotto:2003yf}).
Holes lie within the Fermi sea instead of being separated from it,
and so all the questions as to when and whether there is
an open string description apply here as well.
The worldsheet description of holes is an important missing entry in 
our translation table.

It is interesting that the open/closed string equivalence
in this system is built out of objects that don't 
carry conserved charge,
as opposed to standard examples like D3 branes 
providing the gauge theory dual to $AdS_5\times \IS^5$,
which are charged sources for  
antisymmetric tensors $C_{\sst \!(r)}$.
It raises the question of whether there are other
situations in string theory where there is an
open-closed string equivalence in terms of uncharged
objects.  A good part of the program to understand 
open string tachyon condensation is driven by this question.
Is the late-time dynamics of the open string tachyon
condensate on unstable branes 
(sometimes called {\it tachyon matter})
an alternate description 
of (at least a self-contained subsector of) closed string dynamics?
We have one system where the answer is yes, and it would
be interesting to know if there are others, and if so
whether such an equivalence holds generically
(\cf\ \cite{McGreevy:2003kb} for a discussion
in the present context).


\subsection{Closed string tachyons}

Although the linear dilaton lifts the mass shell of 
the ``tachyon'' to zero in 2D spacetime, the spacelike
tachyon condensate of 2D string theory may still contain
clues to the properties of closed string tachyons
in string theory.  While much of the physics of open string
tachyon condensation is relatively well understood by now,
closed string tachyon condensation is still rather mysterious.
The only controlled examples which have been studied
involve closed string tachyons on localized defects
\cite{Adams:2001sv,Vafa:2001ra,Harvey:2001wm,David:2001vm}
(for reviews, see
\cite{Martinec:2002tz,Headrick:2004hz}).
In these cases, the localized defect decays to flat spacetime
with a pulse of radiation, much like the decay of D-branes
via open string tachyon condensation.
The condensation of delocalized tachyons is less well understood.
The resulting backgrounds will have a cosmological character
since the spacetime geometry will react to the 
stress-energy density of the evolving tachyon field.

Examples of this sort are just beginning to be studied
in 2D string theory.  In a sense, the closed string tachyon
condensate is really only a stationary rather than a static
background of the continuum theory.  From the open string 
point of view, the custodian of this 2D cosmos must sit 
with a bucket of eigenvalues and keep throwing them in at
a constant rate in order to preserve the Fermi sea.
If this entity tires of its task, the Fermi sea drains away;
the corresponding closed string background is then a time-dependent
tachyon field.  Properties of such backgrounds have
been investigated in 
\cite{Alexandrov:2002fh,Karczmarek:2003pv,Karczmarek:2004ph,%
Karczmarek:2004yc,Das:2004hw,Mukhopadhyay:2004ff},
and might serve as a paradigm for the general problem
of closed string tachyon condensates.

\vskip 2cm
\noindent
{{{\bf Acknowledgments}}}:
Thanks to
the organizers of the school for their kind invitation
to present these lectures, 
and to the Aspen Center for Physics
for its hospitality during the assembly of these notes.
I also wish to acknowledge support from DOE grant DE-FG02-90ER-40560.


\bibliographystyle{utphys}
\bibliography{biblio}

\providecommand{\href}[2]{#2}\begingroup\raggedright\begin{thebibliography}{10}

\bibitem{Aharony:1999ti}
O.~Aharony, S.~S. Gubser, J.~M. Maldacena, H.~Ooguri, and Y.~Oz, ``Large n
  field theories, string theory and gravity,'' {\em Phys. Rept.} {\bf 323}
  (2000) 183--386, \href{http://xxx.lanl.gov/abs/hep-th/9905111}{{\tt
  hep-th/9905111}}.

\bibitem{Witten:1998zw}
E.~Witten, ``Anti-de sitter space, thermal phase transition, and confinement in
  gauge theories,'' {\em Adv. Theor. Math. Phys.} {\bf 2} (1998) 505--532,
  \href{http://xxx.lanl.gov/abs/hep-th/9803131}{{\tt hep-th/9803131}}.

\bibitem{Peet:2000hn}
A.~W. Peet, ``Tasi lectures on black holes in string theory,''
  \href{http://xxx.lanl.gov/abs/hep-th/0008241}{{\tt hep-th/0008241}}.

\bibitem{Fidkowski:2003nf}
L.~Fidkowski, V.~Hubeny, M.~Kleban, and S.~Shenker, ``The black hole
  singularity in ads/cft,'' {\em JHEP} {\bf 02} (2004) 014,
  \href{http://xxx.lanl.gov/abs/hep-th/0306170}{{\tt hep-th/0306170}}.

\bibitem{Peet:1998wn}
A.~W. Peet and J.~Polchinski, ``Uv/ir relations in ads dynamics,'' {\em Phys.
  Rev.} {\bf D59} (1999) 065011,
  \href{http://xxx.lanl.gov/abs/hep-th/9809022}{{\tt hep-th/9809022}}.

\bibitem{Banks:1998dd}
T.~Banks, M.~R. Douglas, G.~T. Horowitz, and E.~J. Martinec, ``Ads dynamics
  from conformal field theory,''
  \href{http://xxx.lanl.gov/abs/hep-th/9808016}{{\tt hep-th/9808016}}.

\bibitem{Balasubramanian:1998de}
V.~Balasubramanian, P.~Kraus, A.~E. Lawrence, and S.~P. Trivedi, ``Holographic
  probes of anti-de sitter space-times,'' {\em Phys. Rev.} {\bf D59} (1999)
  104021, \href{http://xxx.lanl.gov/abs/hep-th/9808017}{{\tt hep-th/9808017}}.

\bibitem{Klebanov:1991qa}
I.~R. Klebanov, ``String theory in two-dimensions,''
  \href{http://xxx.lanl.gov/abs/hep-th/9108019}{{\tt hep-th/9108019}}.

\bibitem{Ginsparg:1993is}
P.~H. Ginsparg and G.~W. Moore, ``Lectures on 2-d gravity and 2-d string
  theory,'' \href{http://xxx.lanl.gov/abs/hep-th/9304011}{{\tt
  hep-th/9304011}}.

\bibitem{Dorn:1994xn}
H.~Dorn and H.~J. Otto, ``Two and three point functions in liouville theory,''
  {\em Nucl. Phys.} {\bf B429} (1994) 375--388,
  \href{http://xxx.lanl.gov/abs/hep-th/9403141}{{\tt hep-th/9403141}}.

\bibitem{Zamolodchikov:1995aa}
A.~B. Zamolodchikov and A.~B. Zamolodchikov, ``Structure constants and
  conformal bootstrap in liouville field theory,'' {\em Nucl. Phys.} {\bf B477}
  (1996) 577--605, \href{http://xxx.lanl.gov/abs/hep-th/9506136}{{\tt
  hep-th/9506136}}.

\bibitem{Teschner:1995yf}
J.~Teschner, ``On the liouville three point function,'' {\em Phys. Lett.} {\bf
  B363} (1995) 65--70, \href{http://xxx.lanl.gov/abs/hep-th/9507109}{{\tt
  hep-th/9507109}}.

\bibitem{Fateev:2000ik}
V.~Fateev, A.~B. Zamolodchikov, and A.~B. Zamolodchikov, ``Boundary liouville
  field theory. i: Boundary state and boundary two-point function,''
  \href{http://xxx.lanl.gov/abs/hep-th/0001012}{{\tt hep-th/0001012}}.

\bibitem{Teschner:2000md}
J.~Teschner, ``Remarks on liouville theory with boundary,''
  \href{http://xxx.lanl.gov/abs/hep-th/0009138}{{\tt hep-th/0009138}}.

\bibitem{Zamolodchikov:2001ah}
A.~B. Zamolodchikov and A.~B. Zamolodchikov, ``Liouville field theory on a
  pseudosphere,'' \href{http://xxx.lanl.gov/abs/hep-th/0101152}{{\tt
  hep-th/0101152}}.

\bibitem{Teschner:2001rv}
J.~Teschner, ``Liouville theory revisited,'' {\em Class. Quant. Grav.} {\bf 18}
  (2001) R153--R222, \href{http://xxx.lanl.gov/abs/hep-th/0104158}{{\tt
  hep-th/0104158}}.

\bibitem{Nakayama:2004vk}
Y.~Nakayama, ``Liouville field theory: A decade after the revolution,'' {\em
  Int. J. Mod. Phys.} {\bf A19} (2004) 2771--2930,
  \href{http://xxx.lanl.gov/abs/hep-th/0402009}{{\tt hep-th/0402009}}.

\bibitem{McGreevy:2003kb}
J.~McGreevy and H.~Verlinde, ``Strings from tachyons: The c = 1 matrix
  reloaded,'' {\em JHEP} {\bf 12} (2003) 054,
  \href{http://xxx.lanl.gov/abs/hep-th/0304224}{{\tt hep-th/0304224}}.

\bibitem{Martinec:2003ka}
E.~J. Martinec, ``The annular report on non-critical string theory,''
  \href{http://xxx.lanl.gov/abs/hep-th/0305148}{{\tt hep-th/0305148}}.

\bibitem{Klebanov:2003km}
I.~R. Klebanov, J.~Maldacena, and N.~Seiberg, ``D-brane decay in
  two-dimensional string theory,'' {\em JHEP} {\bf 07} (2003) 045,
  \href{http://xxx.lanl.gov/abs/hep-th/0305159}{{\tt hep-th/0305159}}.

\bibitem{Takayanagi:2003sm}
T.~Takayanagi and N.~Toumbas, ``A matrix model dual of type 0b string theory in
  two dimensions,'' {\em JHEP} {\bf 07} (2003) 064,
  \href{http://xxx.lanl.gov/abs/hep-th/0307083}{{\tt hep-th/0307083}}.

\bibitem{Douglas:2003up}
M.~R. Douglas, I.~R. Klebanov, D.~Kutasov, J.~Maldacena, E.~Martinec, and
  N.~Seiberg, ``A new hat for the c = 1 matrix model,''
  \href{http://xxx.lanl.gov/abs/hep-th/0307195}{{\tt hep-th/0307195}}.

\bibitem{Sen:1999mg}
A.~Sen, ``Non-bps states and branes in string theory,''
  \href{http://xxx.lanl.gov/abs/hep-th/9904207}{{\tt hep-th/9904207}}.

\bibitem{Martinec:2002tz}
E.~J. Martinec, ``Defects, decay, and dissipated states,''
  \href{http://xxx.lanl.gov/abs/hep-th/0210231}{{\tt hep-th/0210231}}.

\bibitem{Green:1987sp}
M.~B. Green, J.~H. Schwarz, and E.~Witten, ``Supestring theory, vols. 1 and
  2,''. Cambridge, Uk: Univ. Pr. (1987) ( Cambridge Monographs On Mathematical
  Physics).

\bibitem{Polchinski:1998rq}
J.~Polchinski, ``String theory. vols. 1 and 2,''. Cambridge, UK: Univ. Pr.
  (1998).

\bibitem{Braaten:1982yn}
E.~Braaten, T.~Curtright, and C.~B. Thorn, ``An exact operator solution of the
  quantum liouville field theory,'' {\em Ann. Phys.} {\bf 147} (1983) 365.

\bibitem{Polyakov:1981rd}
A.~M. Polyakov, ``Quantum geometry of bosonic strings,'' {\em Phys. Lett.} {\bf
  B103} (1981) 207--210.

\bibitem{Knizhnik:1988ak}
V.~G. Knizhnik, A.~M. Polyakov, and A.~B. Zamolodchikov, ``Fractal structure of
  2d-quantum gravity,'' {\em Mod. Phys. Lett.} {\bf A3} (1988) 819.

\bibitem{David:1988hj}
F.~David, ``Conformal field theories coupled to 2-d gravity in the conformal
  gauge,'' {\em Mod. Phys. Lett.} {\bf A3} (1988) 1651.

\bibitem{Distler:1988jt}
J.~Distler and H.~Kawai, ``Conformal field theory and 2-d quantum gravity or
  who's afraid of joseph liouville?,'' {\em Nucl. Phys.} {\bf B321} (1989) 509.

\bibitem{Polchinski:1990mf}
J.~Polchinski, ``Critical behavior of random surfaces in one-dimension,'' {\em
  Nucl. Phys.} {\bf B346} (1990) 253--263.

\bibitem{Brezin:1977sv}
E.~Brezin, C.~Itzykson, G.~Parisi, and J.~B. Zuber, ``Planar diagrams,'' {\em
  Commun. Math. Phys.} {\bf 59} (1978) 35.

\bibitem{Gross:1990md}
D.~J. Gross and I.~R. Klebanov, ``Vortices and the nonsinglet sector of the c =
  1 matrix model,'' {\em Nucl. Phys.} {\bf B354} (1991) 459--474.

\bibitem{Boulatov:1991xz}
D.~Boulatov and V.~Kazakov, ``One-dimensional string theory with vortices as
  the upside down matrix oscillator,'' {\em Int. J. Mod. Phys.} {\bf A8} (1993)
  809--852, \href{http://xxx.lanl.gov/abs/hep-th/0012228}{{\tt
  hep-th/0012228}}.

\bibitem{Das:1990ka}
S.~R. Das and A.~Jevicki, ``String field theory and physical interpretation of
  d = 1 strings,'' {\em Mod. Phys. Lett.} {\bf A5} (1990) 1639--1650.

\bibitem{Polchinski:1991uq}
J.~Polchinski, ``Classical limit of (1+1)-dimensional string theory,'' {\em
  Nucl. Phys.} {\bf B362} (1991) 125--140.

\bibitem{Zinn-Justin:1980uk}
J.~Zinn-Justin, ``Perturbation series at large orders in quantum mechanics and
  field theories: Application to the problem of resummation,'' {\em Phys.
  Rept.} {\bf 70} (1981) 109.

\bibitem{Shenker:1990uf}
S.~H. Shenker, ``The strength of nonperturbative effects in string theory,''.
  Presented at the Cargese Workshop on Random Surfaces, Quantum Gravity and
  Strings, Cargese, France, May 28 - Jun 1, 1990.

\bibitem{Balasubramanian:2001nh}
V.~Balasubramanian, M.~Berkooz, A.~Naqvi, and M.~J. Strassler, ``Giant
  gravitons in conformal field theory,'' {\em JHEP} {\bf 04} (2002) 034,
  \href{http://xxx.lanl.gov/abs/hep-th/0107119}{{\tt hep-th/0107119}}.

\bibitem{Moore:1991zv}
G.~W. Moore, M.~R. Plesser, and S.~Ramgoolam, ``Exact s matrix for 2-d string
  theory,'' {\em Nucl. Phys.} {\bf B377} (1992) 143--190,
  \href{http://xxx.lanl.gov/abs/hep-th/9111035}{{\tt hep-th/9111035}}.

\bibitem{Moore:1992gb}
G.~W. Moore and R.~Plesser, ``Classical scattering in (1+1)-dimensional string
  theory,'' {\em Phys. Rev.} {\bf D46} (1992) 1730--1736,
  \href{http://xxx.lanl.gov/abs/hep-th/9203060}{{\tt hep-th/9203060}}.

\bibitem{Witten:1991zd}
E.~Witten, ``Ground ring of two-dimensional string theory,'' {\em Nucl. Phys.}
  {\bf B373} (1992) 187--213,
  \href{http://xxx.lanl.gov/abs/hep-th/9108004}{{\tt hep-th/9108004}}.

\bibitem{Sen:2004zm}
A.~Sen, ``Rolling tachyon boundary state, conserved charges and two dimensional
  string theory,'' {\em JHEP} {\bf 05} (2004) 076,
  \href{http://xxx.lanl.gov/abs/hep-th/0402157}{{\tt hep-th/0402157}}.

\bibitem{Sen:2004yv}
A.~Sen, ``Symmetries, conserved charges and (black) holes in two dimensional
  string theory,'' \href{http://xxx.lanl.gov/abs/hep-th/0408064}{{\tt
  hep-th/0408064}}.

\bibitem{Moore:1991sf}
G.~W. Moore, ``Double scaled field theory at c = 1,'' {\em Nucl. Phys.} {\bf
  B368} (1992) 557--590.

\bibitem{Moore:1991ir}
G.~W. Moore, N.~Seiberg, and M.~Staudacher, ``From loops to states in 2-d
  quantum gravity,'' {\em Nucl. Phys.} {\bf B362} (1991) 665--709.

\bibitem{Moore:1991ag}
G.~W. Moore and N.~Seiberg, ``From loops to fields in 2-d quantum gravity,''
  {\em Int. J. Mod. Phys.} {\bf A7} (1992) 2601--2634.

\bibitem{Natsuume:1994sp}
M.~Natsuume and J.~Polchinski, ``Gravitational scattering in the c = 1 matrix
  model,'' {\em Nucl. Phys.} {\bf B424} (1994) 137--154,
  \href{http://xxx.lanl.gov/abs/hep-th/9402156}{{\tt hep-th/9402156}}.

\bibitem{DiFrancesco:1991ss}
P.~Di~Francesco and D.~Kutasov, ``Correlation functions in 2-d string theory,''
  {\em Phys. Lett.} {\bf B261} (1991) 385--390.

\bibitem{DiFrancesco:1991ud}
P.~Di~Francesco and D.~Kutasov, ``World sheet and space-time physics in
  two-dimensional (super)string theory,'' {\em Nucl. Phys.} {\bf B375} (1992)
  119--172, \href{http://xxx.lanl.gov/abs/hep-th/9109005}{{\tt
  hep-th/9109005}}.

\bibitem{Bershadsky:1990xb}
M.~Bershadsky and I.~R. Klebanov, ``Genus one path integral in two-dimensional
  quantum gravity,'' {\em Phys. Rev. Lett.} {\bf 65} (1990) 3088--3091.

\bibitem{Sakai:1990ag}
N.~Sakai and Y.~Tanii, ``Compact boson coupled to two-dimensional gravity,''
  {\em Int. J. Mod. Phys.} {\bf A6} (1991) 2743--2754.

\bibitem{Callan:1994ub}
J.~Callan, Curtis~G., I.~R. Klebanov, A.~W.~W. Ludwig, and J.~M. Maldacena,
  ``Exact solution of a boundary conformal field theory,'' {\em Nucl. Phys.}
  {\bf B422} (1994) 417--448,
  \href{http://xxx.lanl.gov/abs/hep-th/9402113}{{\tt hep-th/9402113}}.

\bibitem{Kutasov:2004fg}
D.~Kutasov, K.~Okuyama, J.~Park, N.~Seiberg, and D.~Shih, ``Annulus amplitudes
  and zz branes in minimal string theory,'' {\em JHEP} {\bf 08} (2004) 026,
  \href{http://xxx.lanl.gov/abs/hep-th/0406030}{{\tt hep-th/0406030}}.

\bibitem{Dhar:1995gw}
A.~Dhar, G.~Mandal, and S.~R. Wadia, ``Discrete state moduli of string theory
  from the c=1 matrix model,'' {\em Nucl. Phys.} {\bf B454} (1995) 541--560,
  \href{http://xxx.lanl.gov/abs/hep-th/9507041}{{\tt hep-th/9507041}}.

\bibitem{Friedan:1985ge}
D.~Friedan, E.~J. Martinec, and S.~H. Shenker, ``Conformal invariance,
  supersymmetry and string theory,'' {\em Nucl. Phys.} {\bf B271} (1986) 93.

\bibitem{Fukuda:2002bv}
T.~Fukuda and K.~Hosomichi, ``Super liouville theory with boundary,'' {\em
  Nucl. Phys.} {\bf B635} (2002) 215--254,
  \href{http://xxx.lanl.gov/abs/hep-th/0202032}{{\tt hep-th/0202032}}.

\bibitem{Ahn:2002ev}
C.~Ahn, C.~Rim, and M.~Stanishkov, ``Exact one-point function of n = 1
  super-liouville theory with boundary,'' {\em Nucl. Phys.} {\bf B636} (2002)
  497--513, \href{http://xxx.lanl.gov/abs/hep-th/0202043}{{\tt
  hep-th/0202043}}.

\bibitem{Gaberdiel:2001zq}
M.~R. Gaberdiel and A.~Recknagel, ``Conformal boundary states for free bosons
  and fermions,'' {\em JHEP} {\bf 11} (2001) 016,
  \href{http://xxx.lanl.gov/abs/hep-th/0108238}{{\tt hep-th/0108238}}.

\bibitem{Takayanagi:2004jz}
T.~Takayanagi, ``Notes on d-branes in 2d type 0 string theory,'' {\em JHEP}
  {\bf 05} (2004) 063, \href{http://xxx.lanl.gov/abs/hep-th/0402196}{{\tt
  hep-th/0402196}}.

\bibitem{Witten:1998cd}
E.~Witten, ``D-branes and k-theory,'' {\em JHEP} {\bf 12} (1998) 019,
  \href{http://xxx.lanl.gov/abs/hep-th/9810188}{{\tt hep-th/9810188}}.

\bibitem{Sen:1999mh}
A.~Sen, ``Descent relations among bosonic d-branes,'' {\em Int. J. Mod. Phys.}
  {\bf A14} (1999) 4061--4078,
  \href{http://xxx.lanl.gov/abs/hep-th/9902105}{{\tt hep-th/9902105}}.

\bibitem{Harvey:2000tv}
J.~A. Harvey and P.~Kraus, ``D-branes as unstable lumps in bosonic open string
  field theory,'' {\em JHEP} {\bf 04} (2000) 012,
  \href{http://xxx.lanl.gov/abs/hep-th/0002117}{{\tt hep-th/0002117}}.

\bibitem{Harvey:2000jt}
J.~A. Harvey, P.~Kraus, F.~Larsen, and E.~J. Martinec, ``D-branes and strings
  as non-commutative solitons,'' {\em JHEP} {\bf 07} (2000) 042,
  \href{http://xxx.lanl.gov/abs/hep-th/0005031}{{\tt hep-th/0005031}}.

\bibitem{Sen:2002nu}
A.~Sen, ``Rolling tachyon,'' {\em JHEP} {\bf 04} (2002) 048,
  \href{http://xxx.lanl.gov/abs/hep-th/0203211}{{\tt hep-th/0203211}}.

\bibitem{Lian:1991ty}
B.~H. Lian and G.~J. Zuckerman, ``2-d gravity with c = 1 matter,'' {\em Phys.
  Lett.} {\bf B266} (1991) 21--28.

\bibitem{Bouwknegt:1991yg}
P.~Bouwknegt, J.~G. McCarthy, and K.~Pilch, ``Brst analysis of physical states
  for 2-d gravity coupled to c <= 1 matter,'' {\em Commun. Math. Phys.} {\bf
  145} (1992) 541--560.

\bibitem{Witten:1992yj}
E.~Witten and B.~Zwiebach, ``Algebraic structures and differential geometry in
  $2-d$ string theory,'' {\em Nucl. Phys.} {\bf B377} (1992) 55--112,
  \href{http://xxx.lanl.gov/abs/hep-th/9201056}{{\tt hep-th/9201056}}.

\bibitem{Banks:1991sg}
T.~Banks, ``The tachyon potential in string theory,'' {\em Nucl. Phys.} {\bf
  B361} (1991) 166--172.

\bibitem{Tseytlin:1991bu}
A.~A. Tseytlin, ``On the tachyonic terms in the string effective action,'' {\em
  Phys. Lett.} {\bf B264} (1991) 311--318.

\bibitem{Witten:1991yr}
E.~Witten, ``On string theory and black holes,'' {\em Phys. Rev.} {\bf D44}
  (1991) 314--324.

\bibitem{Mandal:1991tz}
G.~Mandal, A.~M. Sengupta, and S.~R. Wadia, ``Classical solutions of
  two-dimensional string theory,'' {\em Mod. Phys. Lett.} {\bf A6} (1991)
  1685--1692.

\bibitem{Gibbons:1992rh}
G.~W. Gibbons and M.~J. Perry, ``The physics of 2-d stringy space-times,'' {\em
  Int. J. Mod. Phys.} {\bf D1} (1992) 335--354,
  \href{http://xxx.lanl.gov/abs/hep-th/9204090}{{\tt hep-th/9204090}}.

\bibitem{Nappi:1992as}
C.~R. Nappi and A.~Pasquinucci, ``Thermodynamics of two-dimensional black
  holes,'' {\em Mod. Phys. Lett.} {\bf A7} (1992) 3337--3346,
  \href{http://xxx.lanl.gov/abs/gr-qc/9208002}{{\tt gr-qc/9208002}}.

\bibitem{Teschner:1997ft}
J.~Teschner, ``On structure constants and fusion rules in the sl(2,c)/su(2)
  wznw model,'' {\em Nucl. Phys.} {\bf B546} (1999) 390--422,
  \href{http://xxx.lanl.gov/abs/hep-th/9712256}{{\tt hep-th/9712256}}.

\bibitem{Giveon:2001up}
A.~Giveon and D.~Kutasov, ``Notes on ads(3),'' {\em Nucl. Phys.} {\bf B621}
  (2002) 303--336, \href{http://xxx.lanl.gov/abs/hep-th/0106004}{{\tt
  hep-th/0106004}}.

\bibitem{FZZduality}
V.~Fateev, A.~B. Zamolodchikov, and A.~B. Zamolodchikov, ``unpublished,''.

\bibitem{Horowitz:1996nw}
G.~T. Horowitz and J.~Polchinski, ``A correspondence principle for black holes
  and strings,'' {\em Phys. Rev.} {\bf D55} (1997) 6189--6197,
  \href{http://xxx.lanl.gov/abs/hep-th/9612146}{{\tt hep-th/9612146}}.

\bibitem{Kazakov:2000pm}
V.~Kazakov, I.~K. Kostov, and D.~Kutasov, ``A matrix model for the
  two-dimensional black hole,'' {\em Nucl. Phys.} {\bf B622} (2002) 141--188,
  \href{http://xxx.lanl.gov/abs/hep-th/0101011}{{\tt hep-th/0101011}}.

\bibitem{Alexandrov:2002fh}
S.~Y. Alexandrov, V.~A. Kazakov, and I.~K. Kostov, ``Time-dependent backgrounds
  of 2d string theory,'' {\em Nucl. Phys.} {\bf B640} (2002) 119--144,
  \href{http://xxx.lanl.gov/abs/hep-th/0205079}{{\tt hep-th/0205079}}.

\bibitem{Karczmarek:2003pv}
J.~L. Karczmarek and A.~Strominger, ``Matrix cosmology,'' {\em JHEP} {\bf 04}
  (2004) 055, \href{http://xxx.lanl.gov/abs/hep-th/0309138}{{\tt
  hep-th/0309138}}.

\bibitem{Martinec:2004qt}
E.~Martinec and K.~Okuyama, ``Scattered results in 2d string theory,''
  \href{http://xxx.lanl.gov/abs/hep-th/0407136}{{\tt hep-th/0407136}}.

\bibitem{Tada:1991pa}
T.~Tada, ``(q,p) critical point from two matrix models,'' {\em Phys. Lett.}
  {\bf B259} (1991) 442--447.

\bibitem{Daul:1993bg}
J.~M. Daul, V.~A. Kazakov, and I.~K. Kostov, ``Rational theories of 2-d gravity
  from the two matrix model,'' {\em Nucl. Phys.} {\bf B409} (1993) 311--338,
  \href{http://xxx.lanl.gov/abs/hep-th/9303093}{{\tt hep-th/9303093}}.

\bibitem{Kostov:1991cg}
I.~K. Kostov, ``Strings with discrete target space,'' {\em Nucl. Phys.} {\bf
  B376} (1992) 539--598, \href{http://xxx.lanl.gov/abs/hep-th/9112059}{{\tt
  hep-th/9112059}}.

\bibitem{Okuda:2002yd}
T.~Okuda and S.~Sugimoto, ``Coupling of rolling tachyon to closed strings,''
  {\em Nucl. Phys.} {\bf B647} (2002) 101--116,
  \href{http://xxx.lanl.gov/abs/hep-th/0208196}{{\tt hep-th/0208196}}.

\bibitem{Gaiotto:2003yf}
D.~Gaiotto, N.~Itzhaki, and L.~Rastelli, ``On the bcft description of holes in
  the c = 1 matrix model,'' {\em Phys. Lett.} {\bf B575} (2003) 111--114,
  \href{http://xxx.lanl.gov/abs/hep-th/0307221}{{\tt hep-th/0307221}}.

\bibitem{Adams:2001sv}
A.~Adams, J.~Polchinski, and E.~Silverstein, ``Don't panic! closed string
  tachyons in ale space-times,'' {\em JHEP} {\bf 10} (2001) 029,
  \href{http://xxx.lanl.gov/abs/hep-th/0108075}{{\tt hep-th/0108075}}.

\bibitem{Vafa:2001ra}
C.~Vafa, ``Mirror symmetry and closed string tachyon condensation,''
  \href{http://xxx.lanl.gov/abs/hep-th/0111051}{{\tt hep-th/0111051}}.

\bibitem{Harvey:2001wm}
J.~A. Harvey, D.~Kutasov, E.~J. Martinec, and G.~Moore, ``Localized tachyons
  and rg flows,'' \href{http://xxx.lanl.gov/abs/hep-th/0111154}{{\tt
  hep-th/0111154}}.

\bibitem{David:2001vm}
J.~R. David, M.~Gutperle, M.~Headrick, and S.~Minwalla, ``Closed string tachyon
  condensation on twisted circles,'' {\em JHEP} {\bf 02} (2002) 041,
  \href{http://xxx.lanl.gov/abs/hep-th/0111212}{{\tt hep-th/0111212}}.

\bibitem{Headrick:2004hz}
M.~Headrick, S.~Minwalla, and T.~Takayanagi, ``Closed string tachyon
  condensation: An overview,'' {\em Class. Quant. Grav.} {\bf 21} (2004)
  S1539--S1565, \href{http://xxx.lanl.gov/abs/hep-th/0405064}{{\tt
  hep-th/0405064}}.

\bibitem{Karczmarek:2004ph}
J.~L. Karczmarek and A.~Strominger, ``Closed string tachyon condensation at c =
  1,'' {\em JHEP} {\bf 05} (2004) 062,
  \href{http://xxx.lanl.gov/abs/hep-th/0403169}{{\tt hep-th/0403169}}.

\bibitem{Karczmarek:2004yc}
J.~L. Karczmarek, A.~Maloney, and A.~Strominger, ``Hartle-hawking vacuum for c
  = 1 tachyon condensation,''
  \href{http://xxx.lanl.gov/abs/hep-th/0405092}{{\tt hep-th/0405092}}.

\bibitem{Das:2004hw}
S.~R. Das, J.~L. Davis, F.~Larsen, and P.~Mukhopadhyay, ``Particle production
  in matrix cosmology,'' {\em Phys. Rev.} {\bf D70} (2004) 044017,
  \href{http://xxx.lanl.gov/abs/hep-th/0403275}{{\tt hep-th/0403275}}.

\bibitem{Mukhopadhyay:2004ff}
P.~Mukhopadhyay, ``On the problem of particle production in c = 1 matrix
  model,'' {\em JHEP} {\bf 08} (2004) 032,
  \href{http://xxx.lanl.gov/abs/hep-th/0406029}{{\tt hep-th/0406029}}.

\end{thebibliography}\endgroup


\end{document}